\documentclass[preprint2]{proto}
\usepackage{times, graphicx}
\usepackage{natbib}

\newcommand{\refs}{\par\noindent\hangindent=1pc\hangafter=1}
\begin{document}

\title{\textbf{\LARGE The Long-Term Dynamical Evolution of Planetary Systems}}

\author {\textbf{\large Melvyn B. Davies}}
\affil{\small\em Lund University}
\author {\textbf{\large Fred C. Adams}}
\affil{\small\em University of Michigan}
\author {\textbf{\large Philip  Armitage}}
\affil{\small\em University of Colorado, Boulder}

\author {\textbf{\large John Chambers}}
\affil{\small\em Carnegie Institution of Washington}
\author {\textbf{\large Eric Ford}}
\affil{\small\em The Pennsylvania State University, University of Florida}
\author {\textbf{\large Alessandro Morbidelli}}
\affil{\small\em University of Nice}

\author {\textbf{\large Sean N. Raymond}}
\affil{\small\em University of Bordeaux}

\author {\textbf{\large Dimitri Veras}}
\affil{\small\em University of Cambridge}

\begin{abstract}
\baselineskip = 11pt
\leftskip = 0.65in 
\rightskip = 0.65in
\parindent=1pc
{\small This chapter concerns the long-term dynamical evolution of planetary
systems from both theoretical and observational perspectives. We begin
by discussing the planet-planet interactions that take place within
our own Solar System. We then describe such interactions in more
tightly-packed planetary systems. As planet-planet interactions build
up, some systems become dynamically unstable, leading to strong
encounters and ultimately either ejections or collisions of planets.
After discussing the basic physical processes involved, we consider
how these interactions apply to extrasolar planetary systems and
explore the constraints provided by observed systems. The presence of
a residual planetesimal disc can lead to planetary migration and hence
cause instabilities induced by resonance crossing; however, such discs
can also stabilise planetary systems. The crowded birth environment of
a planetary system can have a significant impact: close encounters and
binary companions can act to destabilise systems, or sculpt their
properties. In the case of binaries, the Kozai mechanism can place
planets on extremely eccentric orbits which may later circularise to
produce hot Jupiters.
 \\~\\~\\~}

\end{abstract}  

\bigskip
\section{\textbf{INTRODUCTION}}

Currently observed planetary systems have typically evolved  between 
the time when the last gas in the protoplanetary disc was dispersed, and today. The clearest 
evidence for this assertion comes from the distribution of Kuiper belt objects in the 
outer solar system, and from the eccentricities of massive extrasolar planets, but many other 
observed properties of planetary systems may also plausibly be the consequence of dynamical 
evolution. This chapter summarizes the different types of gravitational 
interactions that lead to long-term evolution of planetary systems, and reviews the 
application of theoretical models to observations of the solar system and extrasolar planetary 
systems.

Planetary systems evolve due to the exchange of angular momentum and / or energy 
among multiple planets, between planets and disks of numerous small bodies (``planetesimals"), 
between planets and other stars, and via tides with the stellar host. A diverse 
array of dynamical evolution ensues. In the simplest cases, such as a well-separated 
two planet system, the mutual perturbations lead only to periodic oscillations in the 
planets' eccentricity and inclination. Of greater interest are more complex multiple planet systems where 
the dynamics is chaotic. In different circumstances the chaos can lead to unpredictable 
(but bounded) excursions in planetary orbits, to large increases in eccentricity as the 
system explores the full region of phase space allowed by conservation laws, or to 
close approaches between planets resulting in collisions or ejections. Qualitative 
changes to the architecture of planetary systems can likewise be caused by dynamical 
interactions in binary systems, by stellar encounters in clusters, or by changes to 
planetary orbits due to interactions with planetesimal discs.

Theoretically, there has been substantial progress since the last {\em Protostars and Planets} 
meeting in understanding the dynamics that can reshape planetary systems. Observational 
progress has been yet more dramatic. Radial velocity surveys and the Kepler mission have 
provided extensive catalogues of single and multiple planet systems, that can be used to 
constrain the prior dynamical evolution of planetary systems
(see the chapter by {\it Fischer et al.} for more details).
Routine measurements of 
the Rossiter-McLaughlin effect for transiting extrasolar planets have shown that a significant 
fraction of hot Jupiters have orbits that are misaligned with respect to the stellar rotation 
axis, and have prompted new models for how hot Jupiters form. Despite this wealth of 
data, the relative importance of different dynamical processes in producing what we see 
remains unclear, and we will discuss in this review what new data is needed to break 
degeneracies in the predictions of theoretical models. Also uncertain is which observed 
properties of planetary systems reflect dynamical evolution taking place subsequent 
to the dispersal of the gas disk (the subject of this chapter), and which involve the 
coupled dynamics and {\em hydrodynamics}  of planets, planetesimals and gas 
within the protoplanetary disc. The chapter by {\em Baruteau et al}. (2013) reviews this earlier phase of evolution.

We begin this chapter by considering the long term stability of the solar 
system. The solar system is chaotic, but our four giant planets are fundamentally stable, and 
there is only a small probability that the terrestrial planets will experience instability during the 
remaining main-sequence lifetime of the Sun. We then compare the current solar system 
to more tightly-packed planetary systems, which are hypothesized progenitors to both the 
solar system and extrasolar planetary systems. We discuss the conditions, time scales and 
outcomes of the dynamical instabilities that can be present in such systems, and compare 
theoretical models to the observed population of extrasolar planets. We then review how 
interactions between planets and residual planetesimal disks can lead to planetary 
migration, which depending on the circumstances can either stabilize or destabilize 
a planetary system. Finally we discuss the outcome of dynamical interactions between 
planetary systems and other stars, whether bound in binaries or interlopers that perturb 
planets around stars in stellar clusters. Dynamical evolution driven by inclined stellar-mass 
(and possibly substellar or planetary-mass) companions provides a route to the formation of 
hot Jupiters whose orbits are misaligned to the stellar equator, and we review the status of 
models for this process (often called the Kozai mechanism).  We close with a summary
of the key points of this chapter.

\bigskip
\section{\textbf{THE SOLAR SYSTEM TODAY}}

A quick glance at our system, with the planets moving on
quasi-circular and almost coplanar orbits, well separated from each
other, suggests the idea of a perfect clockwork system, where the
orbital frequencies tick the time with unsurpassable precision. But is
it really so? In reality, due to their mutual perturbations, the
orbits of the planets must vary over time.

To a first approximation, these variations can be described by a secular theory developed by Lagrange and Laplace 
\citep[see][]{murray99a}
in which the orbital elements that describe a fixed Keplerian orbit change slowly over time. The variations can be found using Hamilton's equations, expanding the Hamiltonian in a power series in terms of the eccentricity $e$ and inclination $i$ of each planet, and neglecting high-frequency terms that depend on the mean longitudes. Only the lowest order terms are retained since $e$ and $i$ are small for the planetary orbits. The variations for a system of planets $j$ (ranging from 1 to $N$) can then be expressed as
\begin{eqnarray}
e_j\sin\varpi_j&=&\sum_{k=1}^Ne_{kj}\sin(g_kt+\beta_k) \nonumber \\
e_j\cos\varpi_j&=&\sum_{k=1}^Ne_{kj}\cos(g_kt+\beta_k)
\label{daacfmrv_equation1}
\end{eqnarray}
with similar expressions for $i$. Here $\varpi$ is the longitude of perihelion, and the quantities $e_{kj}$, $g_k$, and $\beta_k$ are determined by the planet's masses and initial orbits.

In the Lagrange-Laplace theory, the orbits' semi-major axes $a$ remain
constant, while $e$ and $i$ undergo oscillations with periods of
hundreds of thousands of years. The orbits change, but the variations
are bounded, and there are no long-term trends. Even at peak values,
the eccentricities are small enough that the orbits do not come close
to intersecting. Therefore, the Lagrange-Laplace theory concludes that
the solar system is stable.

The reality, however, is not so simple. The Lagrange-Laplace theory
has several drawbacks that limit its usefulness in real planetary
systems. It is restricted to small values of $e$ and $i$; theories
based on higher order expansions exist, but they describe a much more
complex time-dependence of eccentricities and inclinations, whose
Fourier expansions involve harmonics with argument $\nu t$ where
$\nu=\sum_{k=1}^N n_k g_k + m_k s_k$ and $n_k, m_k$ are integers, and
$g$ and $s$ are secular frequencies associated with $e$ and $i$
respectively. The coefficients of these harmonics are roughly
inversely proportional to $\nu$, so that the Fourier Series
representation breaks down when $\nu\sim 0$, a situation called 
{\it secular resonance}.

Moreover, the Lagrange-Laplace theory ignores the effects of
mean-motion resonances or near resonances between the orbital periods
of the planets. The existence of mean-motion resonances can
fundamentally change the dynamics of a planetary system and alter its
stability in ways not predicted by Lagrange-Laplace theory. In
particular, the terms dependent on the orbital frequencies, ignored in
the Lagrange-Laplace theory, become important when the ratio of two
orbital periods is close to the ratio of two integers. This situation
arises whenever the critical argument $\phi$ varies slowly over time,
where
\begin{equation}
\phi=k_1\lambda_i+k_2\lambda_j+k_3\varpi_i+k_4\varpi_j+k_5\Omega_i+k_6\Omega_j
\label{daacfmrv_equation2}
\end{equation}
for planets $i$ and $j$, where $\lambda$ is the mean longitude,
$\Omega$ is the longitude of the ascending node, and $k_{1-6}$ are
integers. The $k_{3-6}$ terms are included because the orbits will
precess in general, so a precise resonance is slightly displaced from
the case where $k_1/k_2$ is a ratio of integers.

To leading order, the evolution at a resonance can be described quite
well using the equation of motion for a pendulum
\citep{murray99a}:
\begin{equation}
\ddot{\phi}=-\omega^2\sin\phi
\label{daacfmrv_equation3}
\end{equation}

At the centre of the resonance, $\phi$ and $\dot\phi$ are zero and
remain fixed. When $\phi$ is slightly non-zero initially, $\phi$
librates about the equilibrium point with a frequency $\omega$ that
depends on the amplitude of the librations. The semi-major axes and
eccentricities of the bodies involved in the resonance undergo
oscillations with the same frequency. There is a maximum libration
amplitude for which $\phi$ approaches $\pm\pi$. At larger separations
from the equilibrium point, $\phi$ circulates instead and the system
is no longer bound in resonance.

Whereas a single resonance exhibits a well-behaved, pendulum-like
motion as described above, when multiple resonances overlap the
behavior near the boundary of each resonance becomes erratic, which
leads to chaotic evolution.  Because the frequencies of the angles
$\varpi_{i,j}$ and $\Omega_{i,j}$ are small relative to the orbital
frequencies (i.e. the frequencies of $\lambda_{i,j}$), in general
resonances with the same $k_1, k_2$ but different values of
$k_{3,\ldots,6}$ overlap with each other. Thus, quite generically, a
region of chaotic motion can be found associated with each mean-motion
resonance.

There are no mean-motion resonances between pairs of Solar System
planets. Jupiter and Saturn, however, are close to the 2:5 resonance,
and Uranus and Neptune are close to the 1:2 resonance. In both cases,
the resonant angle $\phi$ is in circulation and does not exhibit
chaotic motion. In general, systems can support $N$-body resonances,
which involve integer combinations of the orbital frequencies of $N$
planets, with $N>2$. The forest of possible resonances becomes rapidly
dense, and increases with the number $N$ of planets involved. As a
result, analytic models become inappropriate to describe precisely the
dynamical evolution of a system with many planets.

Fortunately, modern computers allow the long-term evolution of the
Solar System to be studied numerically using $N$-body integrations.
These calculations can include all relevant gravitational interactions
and avoid the approximations inherent in analytic theories. One
drawback is that $N$-body integrations can never prove the stability
of a system, only its stability for the finite length of an
integration.

$N$-body integrations can be used to distinguish between regular and
chaotic regions, and quantify the strength of chaos, by calculating
the system's Lyapunov exponent $\Gamma$, given by
\begin{equation}
\Gamma=\lim_{t\rightarrow\infty}\frac{\ln[d(t)/d(0)]}{t}
\label{daacfmrv_equation4}
\end{equation}
where $d$ is the separation between two initially neighboring
orbits. Regular orbits diverge from one another at a rate that is a
power of time. Chaotic orbits diverge exponentially over long
timespans, although they can be ``sticky'', mimicking regular motion
for extended time intervals. If the Solar System is chaotic, even a
tiny uncertainty in the current orbits of the planets will make it
impossible to predict their future evolution indefinitely.

One of the first indications that the Solar System is chaotic came
from an 845 Myr integration of the orbits of the outer planets plus
Pluto \citep{sussman88}. Pluto was still considered a planet at the
time due to its grossly overestimated mass. This work showed that
Pluto's orbit is chaotic with a Lyapunov timescale ($T_L=1/\Gamma$) of
20 My. Using a 200 Myr integration of the secular equations of motion,
\citet{laskar89} showed that the 8 major planets are chaotic with a
Lyapunov time of 5 My. This result was later confirmed using full
$N$-body integrations \citep{sussman92}. The source of the chaos is
due to the existence of two secular resonances, with frequencies
$\nu_1=2(g_4 - g_3) - (s_4 - s_3)$ and $\nu_2= (g_1 - g_5) - (s_1 -
s_2)$, i.e. frequencies not appearing in the Lagrange-Laplace theory
\citep{laskar90}.

The four giant planets by themselves may be chaotic \citep{sussman92},
due to a three-body resonance involving Jupiter, Saturn and Uranus
\citep{murray99a}. Assuming that the evolution can be described as a
random diffusion through the chaotic phase space within this
resonance, \citet{murray99a} estimated that it will take $\sim
10^{18}$ y for Uranus's orbit to cross those of its neighbors.
However, careful examination suggests that chaotic and regular
solutions both exist within the current range of uncertainty for the
orbits of the outer planets \citep{guzzo05,hayes08}, so the lifetime
of this subsystem could be longer, and even infinite.

Numerical integrations can also be used to assess the long-term
stability of the planetary system, with the caveat that the precise
evolution can never be known, so that simulations provide only a
statistical measure of the likely behavior. Numerical simulations
confirm the expectation that the orbits of the giant planets will not
change significantly over the lifetime of the Sun \citep{batygin08}.
Nonetheless, there remains a remote possibility that the orbits of the
inner planets will become crossing on a timescale of a few Gyr
\citep{laskar94, laskar08, batygin08, laskar09}.

In principle, if the terrestrial planets were alone in the solar
system, they would be stable for all time. (In practice, of course,
the Solar System architecture will change on a time scale of only
$\sim7$ Gyr due to solar evolution and mass loss.)  In fact, even if
the orbits of the inner planets are free to diffuse through phase
space, their evolution would still be constrained by their total
energy and angular momentum.

A useful ingredient in orbital evolution is the angular momentum
deficit (AMD), given by 
\begin{equation}
 {\rm AMD} \equiv \sum_k \Lambda_k \left( 1 - \cos i_k \sqrt{1 - e_k^2} \right)
\label{daacfmrv_equation5}
\end{equation}
where $\Lambda_k = M_k M_* / (M_k + M_*) \sqrt{G (M_* + M_k) a_k}$ is
the angular momentum of planet $k$ with mass $M_k$, semi-major axis
$a_k$, eccentricity $e_k$ and inclination $i_k$ to the invariable
plane, and $M_*$ is the mass of the host star.

In absence of mean-motion resonances, to a good approximation, $a$
remains constant for each planet, so AMD is also constant
\citep[e.g.][though this result dates to Laplace]{laskar97},
Excursions in $e$ and $i$ are constrained by conservation of AMD, and
the maximum values attainable by any one planet occur when all the
others have $e=i=0$. Mercury's low mass means it can acquire a more
eccentric and inclined orbit than the other planets. However, even
when Mercury absorbs all available AMD, its orbit does not cross
Venus. However, when the giant planets are taken into account, the
exchange of a small amount of angular momentum between the outer
planets and the inner planets can give the latter enough AMD to
develop mutual crossing orbits. As a result, long-term stability is
not guaranteed.

The possible instability of the terrestrial planets on a timescale of
a few Gyr shows that the solar system has not yet finished evolving
dynamically. Its overall structure can still change in the future,
even if marginally within the remaining main-sequence lifetime of the
Sun.

This potential change in the orbital structure of the planets would
not be the first one in the history of the solar system. In fact,
several aspects of the current orbital architecture of the solar
system suggest that the planetary orbits changed quite drastically
after the epoch of planet formation and the disappearence of the
proto-planetary disk of gas. For instance, as described above, the
giant planets are not in mean motion resonance with each other.
However, their early interaction with the disk of gas in which they
formed could have driven them into mutual mean-motion resonances such
as the 1:2, 2:3 or 3:4, where the orbital separations are much
narrower than the current ones \citep{lee02, kley05, morbidelli07}.
We do indeed observe many resonant configurations among extra-solar
planetary systems, and in numerical simulations of planetary
migration.  Also, the current eccentricities and inclinations of the
giant planets of the solar system, even if smaller than those of
many extra-solar planets, are nevertheless non-zero. In a disk of gas,
without any mean-motion resonant interaction, the damping effects
would have annihilated the eccentricities and inclinations of the
giant planets in a few hundred orbits  \citep{kley06, cresswell07}.
So, some mechanism must have extracted the giant planets from any
original mean-motion resonances and placed them onto their current,
partially eccentric and inclined orbits, after gas removal.

The populations of small bodies of the solar system also attest to
significant changes in the aftermath of gas removal, and possibly
several 100 Myr later. The existence and properties of these
populations thus provide important constraints on our dynamical
history (e.g., see Section 6).  There are three main reservoirs of
small bodies: the asteroid belt between Mars and Jupiter, the Kuiper
belt immediately beyond Neptune and the Oort cloud at the outskirts of
the solar system. At the present epoch, both the asteroid belt and the
Kuiper belt contain only a small fraction of the mass (0.1\% or less)
that is thought to exist in these regions when the large objects that
we observe today formed \citep{kenyon04}.  As a result, the vast
majority of the primordial objects (by number) have been dynamically
removed. Some of these bodies were incorporated into the forming
planets, some were scattered to other locations within the solar
system, and some where ejected. The remaining objects (those that make
up the current populations of the asteroid and Kuiper belts) are
dynamically excited, in the sense that their eccentricities and
inclinations cover the entire range of values allowed by long-term
stability constraints; for instance, the inclinations can be as large
as 40 degrees, much larger than the nearly-co-planar orbits of the
planets themselves. This evidence suggests that some dynamical
mechanism removed more than 99.9\% of the objects and left the
survivors on excited orbits, much different from their original
circular and co-planar ones. Another constaint is provided by the
absence of a correlation between the size of the objects and their
orbital excitation. Presumably, the mechanism that altered the orbits
of these small bodies acted {\it after} the removal of the nebular
gas; otherwise, gas drag, which is notoriously a size-dependent
process, would have imprinted such a correlation. Similarly, the Oort
cloud (the source of long period comets) contains hundreds of billions
of kilometer-size objects, and most trace through orbits with high
eccentricities and inclinations \citep{wiegert99, kaib09}. Since the
Oort cloud extends far beyond the expected size of circumstellar
disks, the population is thought to have been scattered to such large
distances. In the presence of gas, however, these small objects could
not have been scattered out by the giant planets, because gas-drag
would have circularized their orbits just beyond the giant planets
\citep{brasser07}. Thus, presumably the formation of the Oort cloud
post-dates gas removal; this timing is consistent with considerations
of the Solar birth environment (Section 6).

Finally, there is evidence for a surge in the impact rates and/or
impact velocities on the bodies of the inner solar system, including
the terrestrial planets, the Moon, and the asteroid Vesta
\citep{tera74, ryder02, kring02, marchi12a, marchi12b, marchi13}.
This surge seems to have occurred during a time interval ranging from
4.1 to 3.8 Gyr ago, i.e., starting about 400 Myr after planet
formation. This event is often referred to as the ``terminal Lunar
Cataclysm" or the ``Late Heavy Bombardment". If such a cataclysm
really happened \citep[its existence is still debated today; see for
  instance][]{hartmann00}, it suggests that the changes in the
structure of the solar system described above did not happen
immediately after the removal of gas from the early solar nebula, 
but only after a significant delay of several 100 Myr.

All of this discussion provides hints that the mechanisms of orbital
instability discussed in the next sections of this chapter in the
framework of extra-solar planets evolution were probably not foreign
to the solar system. Specifically, a scenario of possible evolution of
the solar system that explains all these aspects will be presented in
Section 5. 


\bigskip
\section{\textbf{INSTABILITIES IN TIGHTLY-PACKED SYSTEMS}}

We will see in this section that the timescale for planetary system to become unstable
is a very sensitive function of planetary separations. Thus the separations of 
planets within a system is an important quantity. 
Unfortunately, the typical separation of planets in 
newly-formed planetary systems is unknown 
observationally and theoretical predictions are also unclear.
 For terrestrial planets 
whose final assembly occurs after gas disk dispersal, the results of \citet{kokubo98} 
suggest that separations of $\simeq 10$ mutual Hill radii, $R_{hill,m}$
 are typical. Where\footnote{This is one definition of the mutual 
Hill radius; there are others in use in the literature.}
\begin{equation}
 R_{hill,m}  \equiv \left( \frac{M_k + M_{k+1}}{3 M_*} \right)^{1/3} \frac{a_k + a_{k+1}}{2}
\label{daacfmrv_equation6}
\end{equation}
Here $M_\ast$ is the stellar mass,
$M_k$ are the planetary masses, and $a_k$ the semi-major axes.
However, terrestrial planets that form more rapidly (i.e. before 
gas disk dispersal)
may be more tightly-packed,
 as may also be the case 
 for giant planets that necessarily form in a dissipative environment. Giant 
planets (or their cores) can migrate due to either planetesimal \citep{levison10} 
or gas disk interactions \citep{kley12}. Hydrodynamic simulations of multiple 
planets interacting with each other and with a surrounding gas disk \citep{moeckel08,marzari10,
moeckel12,lega13} show that resonant, tightly packed or well-separated systems can 
form, but the probabilities for these channels cannot be predicted from first 
principles. Constraints on the dynamics must 
currently be derived from comparison of the predicted end states 
with observed systems.

For giant planets, two types of 
instability provide dynamical paths that may explain the origin of hot Jupiters 
and eccentric giant planets. If the planets start on circular, coplanar orbits, 
{\em planet-planet} scattering results in a combination of physical collisions, 
ejection, and generation of eccentricity. If the system instead forms 
with planets on widely-separated eccentric or inclined orbits, {\em secular chaos} can result 
in the diffusive evolution of eccentricity to high values even in the absence 
of close encounters. 

\bigskip
\noindent
\textbf{3.1 Conditions and time scales of instability}
\medskip

The condition for stability can be analytically derived for 
two planet systems. 
\citet{gladman93}, drawing on results from \citet{marchal82} and others, 
showed that two planet systems with initially circular and coplanar orbits are 
Hill stable for separations $\Delta \equiv (a_2 - a_1) / R_{hill,m} \gtrsim 2\sqrt{3}$. 
Hill stability implies that two planets with at least this separation are analytically guaranteed 
to never experience a close approach. Numerically, it is found that some systems can 
be stable at smaller separations within mean motion resonances, and that the stronger 
condition of Lagrange stability --- which requires that both planets remain bound and 
ordered for  all time --- requires only modestly greater spacing than Hill stability \citep{barnes06b,veras13}.

\begin{figure*}[t]
\centering
\includegraphics[width=170mm]{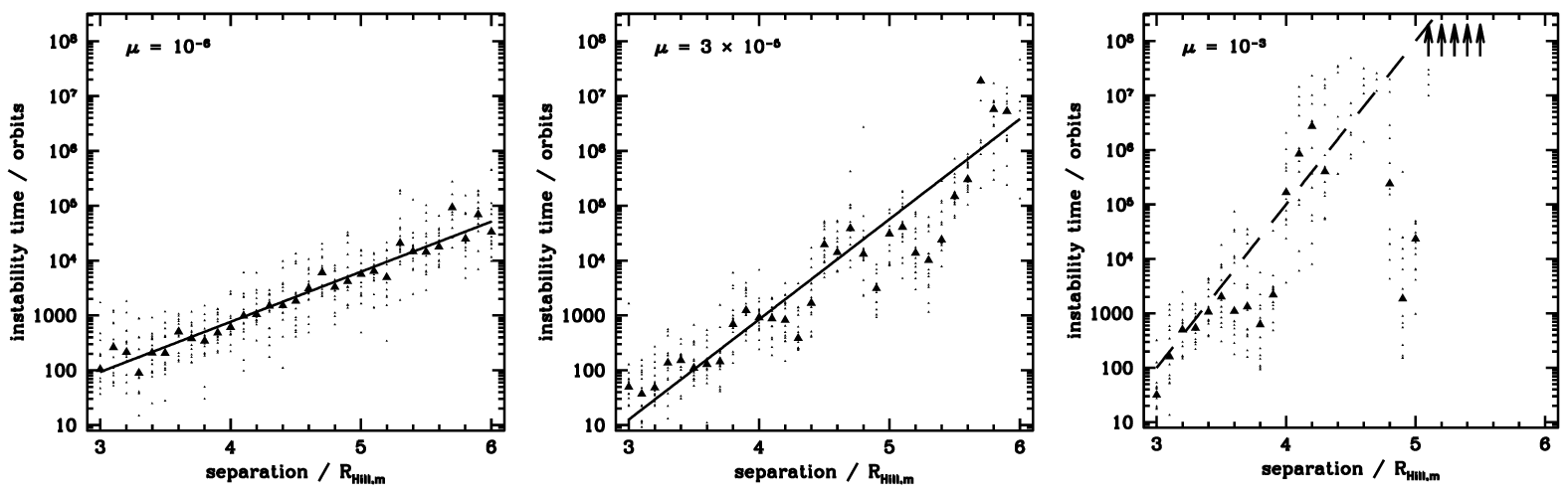}
\caption{The median (triangles) instability time scale for initially circular, coplanar three planet systems, as 
a function of the separation in units of mutual Hill radii \citep[after][]{chambers96,marzari02}. 
The dots show the instability time scale obtained for individual realisations of 
a planetary system. The instability time scale is defined as the time 
(in units of the initial orbital period of the inner planet) until 
the first pair of planets approach within one Hill radius. From left to right, the panels show mass ratios 
$\mu = M_p / M_*$ of $10^{-6}$, $3 \times 10^{-5}$ and $10^{-3}$.}
\label{daacfmrv_figure1}
\end{figure*}

There is no analytic criterion for the absolute stability of systems with $N \ge 3$ planets. The 
degree of instability can be characterized numerically by evaluating the time scale for orbit 
crossing, or for the first close encounters between planets, to occur (in practice, different 
reasonable definitions of ``instability" are nearly equivalent). Figure~\ref{daacfmrv_figure1} 
illustrates the median instability time scale as a function of the separation in units of 
$R_{hill,m}$ for planetary systems 
of three equal mass planets with varying mass ratios $\mu = M_p / M_*$. The behavior is 
simplest for low mass planets. In this regime, \citet{chambers96} found that the time 
before the first close encounters could be approximated 
as $\log(t_{close}) = b \Delta + c$, 
with $b$ and $c$ being constants. 
The time to a first close encounter is found to vary enormously over a small range
of initial planetary separations.
Systems with $N=5$ were less stable 
than the $N=3$ case, but there was little further decrease in the stability time with 
further increase in the planet number to $N=10$ or $N=20$. The scaling of the 
instability time with separation was found to be mass dependent if measured in 
units of $R_{hill,m}$ (with a steeper dependence for larger $\mu$), but approximately 
{\em independent} if measured in units $\propto M_p^{1/4}$. \citet{smith09} extended 
these results with longer integrations. They found that a single slope provided a good 
fit to their data for Earth mass planets for $3.5 \le \Delta \le 8$, but that there was a 
sharp increase in $b$ for $\Delta > 8$. Sufficiently widely separated systems thus rapidly 
become stable for practical purposes. As in the two planet case, resonant systems 
can evade the non-resonant stability scalings, but only for a limited number of 
planets \citep{matsumoto12}.

Similar numerical experiments for more massive planets with $\mu \sim 10^{-3}$ were 
conducted by \citet{marzari02} and by \citet{chatterjee08}. At these mass ratios, the 
plot of $t_{close} (\Delta)$ exhibits a great deal of structure associated with the 
2:1 (and to a lesser extent 3:1) mean-motion resonance (Figure~\ref{daacfmrv_figure1}). 
A simple exponential fit is no longer a good approximation for instability time scales 
of $10^6 - 10^8$~yr (at a few AU) that might be highly relevant for gas giants emerging 
from the gas disk. \citet{chatterjee08} quote an improved fitting formula for the instability 
time (valid away from resonances), but there is an unavoidable dependence on details 
of the system architecture such as the radial ordering of the masses 
in systems with unequal mass planets \citep{raymond10}.

For a single planet, resonance overlap leads to chaos and instability of test particle 
orbits out to a distance that scales with planet mass as $\Delta a \propto \mu^{2/7}$ for low
eccentricities \citep{wisdom80} and $\Delta a \propto \mu^{1/5}$ above a (small)
critical eccentricity \citep{mustill12}.
Resonance overlap is similarly thought to underly the 
numerical results for the stability time of multiple planet systems \citep{lissauer95,morbidelli96}, 
though the details are not fully understood. \citet{quillen11} considered the criterion 
for the overlap of three-body resonances, finding that the density of these resonances was 
(to within an order of magnitude) sufficient to explain the origin of chaos and instability 
at the separations probed numerically by \citet{smith09}.

Once stars depart the main sequence, the increase in $\mu$ as the star loses mass can 
destabilize either multiple planet systems \citep{duncan98,debes02,veras13b,voyatzis13}, 
or asteroid belts 
whose members are driven to encounter mean motion resonances with a single massive 
planet \citep{debes12}. These ``late" instabilities may explain the origin of metal-polluted 
white dwarfs, and white dwarfs with debris disks. 

\bigskip
\noindent
\textbf{3.2 Outcome of instability}
\medskip

The outcome of instability in tightly packed planetary systems is a combination of ejections, 
physical collisions between planets, and planet-star close approaches that may lead to tidal 
dissipation or direct collision with the star. In more widely spaced systems the same 
outcomes can occur, but chaotic motion 
can also persist indefinitely without dramatic dynamical consequences.  
In the inner Solar System, 
for example, the Lyapunov time scale is very short ($\simeq 5$~Myr), but the only known 
pathway to a planetary collision --- via the entry of Mercury into a secular resonance with 
Jupiter --- has a probability of only $P \approx 10^{-2}$ within 5~Gyr \citep{laskar89,batygin08,laskar09}.

 During a close
encounter between two planets, scattering is statistically favored
over collisions when the escape speed from the planets' surfaces is
larger than the escape speed from the planetary
system~\citep{goldreich04}. This can be quantified by the Safronov
number $\Theta$:
\begin{equation}
\Theta^2 = \left(\frac{M_p}{M_\star}\right)  \left(\frac{R_p}{a_p}\right)^{-1}
\label{daacfmrv_equation7}
\end{equation}
where $M_p$, $R_p$ and $a_p$ represent the planet's mass, radius, and
orbital distance, and $M_\star$ is the stellar mass~\citep[see
  discussion in][]{ford08}.  We expect scattering to be most important
among massive, dense planets and in the outer parts of planetary
systems.  In the Solar System, the giant planets' escape speeds are
roughly 2-6 times larger than the highest value for the terrestrial
planets (Earth's).  They are also at larger $a_p$.

Physical collisions lead to 
modest eccentricities for the merged remnants \citep{ford01}. Scattering at orbital 
radii beyond the snow line ($a \approx 3 \ {\rm AU}$), conversely, results in a broad 
eccentricity distribution consistent with that observed for massive extrasolar planets 
\citep{chatterjee08,juric08}. Ejection proceeds via scattering on to highly 
eccentric orbits, and hence a prediction of planet-planet scattering models is the 
existence of a population of planets around young stars with very large orbital 
separation \citep{veras09a,scharf09,malmberg11}.   
The frequency of ejections from scattering is 
probably too small to explain the large abundance of apparently unbound Jupiter-mass objects 
discovered by microlensing \citep{sumi11}, if those are to be free-floating rather than simply on
 wide but bound orbits \citep{veras12}.

The outcome of instabilities that occur within the first few Myr (or as the disk 
disperses) may be modified by gravitational torques or mass accretion from the 
protoplanetary disk, while instabilities toward the outer edges of planetary 
systems will result in interactions with outer planetesimal belts. \citet{matsumura10}, 
using N-body integrations coupled to a one dimensional gas disk model, and 
\citet{moeckel12}, using two dimensional hydrodynamic disk models, found that 
realistic transitions between gas-rich and gas-poor dynamics did not preclude 
the generation of high eccentricities via scattering. Larger numbers of resonant 
systems were, however, predicted. 
\citet{lega13}, instead, found that if a planetary system becomes unstable during the gas-disk phase it is likely to stabilize (after the ejection or the collision of some planets) in resonant low-eccentricity orbits, and avoid further instabilities after that the gas is removed.
\citet{raymond10} ran scattering experiments 
in which planets at larger radii interacted with massive collision-less planetesimal 
disks. The disks strongly suppressed the final eccentricities of lower mass outer 
planetary systems.  

\bigskip
\noindent
\textbf{3.3 Secular chaos}
\medskip

Significantly different evolution is possible in multiple planet systems if one or 
more planets possess  a substantial eccentricity or inclination from the {\em beginning}.
The departure from 
circular coplanar orbits can be quantified by the angular momentum deficit (AMD)
defined earlier.

\citet{wu11} proposed secular chaos (combined with 
tidal effects) as the origin of hot Jupiters. They presented a proof of concept 
numerical integration of a widely separated (but chaotic) planetary system in 
which most of the AMD initially resided in an eccentric ($e \approx 0.3$) outer 
gas giant ($M = 1.5 \ M_J$, $a= 16 \ {\rm AU}$). Diffusion of AMD among the 
three planets in the system eventually led to the innermost planet attaining 
$e \simeq 1$ without prior close encounters among the planets. They noted 
several characteristic features of secular chaos as a mechanism for forming 
hot Jupiters --- it works best for low mass inner planets, requires the presence 
of multiple additional planets at large radii, and can result in star-planet 
tidal interactions that occur very late (in principle after Gyr).

The range of planetary systems for which secular chaos yields dynamically 
interesting outcomes on short enough time scales remains to be quantified. 
A prerequisite is a large enough AMD in the initial conditions. 
This might originate from eccentricity excitation of planets by the gas disk 
\citep[though this is unlikely to occur for low planet masses, e.g.][]{papaloizou01b,dangelo06,dunhill13}, 
from external perturbations such as fly-bys
\citep{malmberg11,boley12}, or from a prior epoch of scattering among 
more tightly packed planets.



\bigskip
\section{\textbf{PLANET-PLANET SCATTERING CONSTRAINED BY 
DYNAMICS AND OBSERVED EXOPLANETS}}

The planet-planet scattering model was developed to explain the
existence of hot Jupiters~\citep{rasio96,weidenschilling96} and giant
planets on very eccentric orbits~\citep{lin97,papaloizou01a,ford01}.
Given the great successes of exoplanet searches 
there now exists a database of observations against which to test the
planet-planet scattering model.
The relevant observational constraints come from the subset of
extra-solar systems containing giant planets.  
In this Section we first review the relevant observational
constraints.  Next, we show how the dynamics of scattering depends on
the parameters of the system.  We then show that, with simple
assumptions, the scattering model can match the observations.

\bigskip
\noindent
\textbf{4.1 Constraints from Giant Exoplanets}
\medskip

The sample of extra-solar planets that can directly constrain models
of planet-planet scattering now numbers more than 300.  These are
giant planets with masses larger than Saturn's and smaller than $13
M_{Jup}$~\citep{wright11,schneider11} with orbital semimajor axes
larger than 0.2 AU to avoid contamination from star-planet tidal
circularization.

The giant planets have a broad eccentricity distribution with a median
of $\sim0.22$~\citep{butler06,udry07}.  There are a number of planets with
very eccentric orbits: roughly 16\% / 6\% / 1\% of the sample has
eccentricities larger than 0.5 / 0.7 / 0.9.

More massive planets have more eccentric orbits.  Giant exoplanets
with minimum masses $M_p > M_{Jup}$ have statistically higher
eccentricities (as measured by a K-S test) than planets with $M_p <
M_{Jup}$~\citep{jones06,ribas07,ford08,wright09}.  Excluding hot
Jupiters, there is no measured correlation between orbital radius and
eccentricity~\citep {ford08}.

Additional constraints can be extracted from multiple planet systems.
For example, the known two-planet systems are observed to cluster just
beyond the Hill stability limit~\citep{barnes06a,raymond09b}.  However,
it's unclear to what extent detection biases contribute to this
clustering. 
In addition, studies of the long-term dynamics of some
well-characterized systems can constrain their secular behavior, and
several systems have been found very close to the boundary between
apsidal libration and circulation~\citep{ford05a,barnes06b,veras09b}.

Often, there are significant uncertainties associated with the
observations~\citep{ford05b}.  Orbital eccentricities are especially
hard to pin down and the current sample may be modestly biased toward
higher eccentricities~\citep{shen08,zakamska11}, although it is clear
that with their near-circular orbits the Solar System's giant planets
are unusual in the context of giant exoplanets.

\bigskip
\noindent
\textbf{4.2 Scattering Experiments: Effect of System Parameters}
\medskip

We now explore how the dynamics and outcomes of planet-planet
scattering are affected by parameters of the giant planets, in
particular their masses and mass ratios.  Our goal is to understand
what initial conditions are able to match all of the observed
constraints from Section 4.1.

During a close gravitational encounter, the magnitude of the
gravitational kick that a planet imparts depends on the planet's
escape speed.  To be more precise, what is important is the Safronov
number $\Theta$, as given in Equation (\ref{daacfmrv_equation7}).  
At a given orbital distance, more massive planets
kick harder, i.e. they impart a stronger change in velocity.  Within
systems with a fixed number of equal-mass planets, more massive
systems evolve more quickly because they require fewer close
encounters to give kicks equivalent to reaching zero orbital energy
and being liberated from the system.  The duration of an instability
-- the time during which the planets' orbits cross -- is thus linked
to the planet masses.  For example, in a set of simulations with three
planets at a few to 10 AU, systems with $M_p = 3 M_{Jup}$ planets
underwent a median of 81 scattering events during the instability, but
this number increased to 175, 542, and 1871 for $M_p = M_{Jup},
M_{Sat}$, and $30M_{\oplus}$, respectively~\citep{raymond10}.  The
instabilities lasted $10^4-10^6$ years; longer for lower-mass planets.

Thus, the timescale for orbital instabilities in a system starts off
longer than the planet formation timescale and decreases as
planets grow their masses.  The natural outcome of giant planet
formation is a planetary system with multiple giant planets which
undergo repeated instabilities on progressively longer timescales,
until the instability time exceeds the age of the system.

The orbits of surviving planets in equal-mass systems also depend on
the planet mass.  Massive planets end up on orbits with larger
eccentricities than less massive
planets~\citep{ford03,raymond08,raymond09a,raymond10}.  This is a
natural consequence of the stronger kicks delivered by more massive
planets.  However, the inclinations of surviving massive planets are
smaller than for less massive planets in terms of both the inclination
with respect to the initial orbital plane (presumably corresponding to
the stellar equator) and the mutual inclination between the orbits of
multiple surviving planets~\citep{raymond10}.  
This increase in
inclination appears to be intimately linked with the number of
encounters the planets have experienced in ejecting other 
planets rather than their strength. 

The planetary mass ratios also play a key role in the scattering
process.  Scattering between equal-mass planets represents the most
energy-intensive scenario for ejection.  Since it requires far less
energy to eject a less massive planet, the recoil that is felt by the
surviving planets is much less.  Thus, the planets that survive
instabilities between unequal-mass planets have smaller eccentricities
and inclinations compared with the planets that survive equal-mass
scattering~\citep{ford03,raymond08,raymond10,raymond12}.  This 
does not appear to be overly sensitive to the planets' ordering,
i.e. if the lower-mass planets are located on interior or exterior
orbits~\citep{raymond10}.

The timing of the instability may also be important for the outcome
because giant planets cool and contract on $10^{7-8}$ year
timescales~\citep{spiegel12}.  Thus, the planets' escape speeds and
$\Theta$ values increase in time.  Instabilities that occur very early
in planetary system histories may thus be less efficient at ejecting
planets and may include a higher rate of collisions compared with most
scattering calculations to date.

\bigskip
\noindent
\textbf{4.3 Scattering Experiments: Matching Observations}
\medskip

Let us consider a simple numerical experiment where all planetary systems
containing giant planets are assumed to form three giant planets.  The
masses of these planets follow the observed mass distribution ${\rm
  d}N/{\rm d}M \propto M^{-1.1}$~\citep{butler06,udry07} and the
masses within a system are not correlated.  All systems become
unstable and undergo planet-planet scattering.  With no fine tuning,
the outcome of this experiment matches the observed eccentricity
distribution~\citep{raymond08}.

The exoplanet eccentricity distribution can be reproduced with a wide
range of initial
conditions~\citep{adams03,moorhead05,juric08,chatterjee08,ford08,raymond10,beauge12}.
The same simulations also reproduce the dynamical quantities that can
be inferred from multiple-planet systems: the distribution of
parameterized distances of two-planet systems from the Hill stability
limit~\citep{raymond09b} and the secular configuration of two-planet
systems~\citep{timpe13}.

Certain observations are difficult to reproduce.  Some observed
systems show no evidence of having undergone an instability (e.g. as
they contain multiple giant planets on near-circular orbits or those in resonances).  The
scattering model must thus be able to reproduce the observed
distributions including a fraction of systems remaining on stable
orbits.  The fraction of systems that are stable  is typically
10-30\%,  where assumptions are made about 
the distribution of planetary masses and orbits ~\citep{juric08,raymond10,raymond11}.
One should also note that migration is required in order
to match the observed distribution of semi-major axes, in other
words scattering alone cannot transport Jupiter-like planets on Jupiter-like
orbits to semi-major axes less than one AU \citep{moorhead05}.

Perhaps the most difficult observation to match is the observed
positive correlation between planet mass and eccentricity.  In the
simulations from the simple experiment mentioned above, lower-mass
surviving planets actually have higher eccentricities than more
massive ones~\citep{raymond10}.  For massive planets to have higher
eccentricities they must form in systems with other massive planets,
since scattering among equal-mass massive planets produces the highest
eccentricities.  This is in agreement with planet formation models:
when the conditions are ripe for giant planet formation (e.g. massive
disk, high metallicity), one would expect all giants to have high
masses, and vice versa.

\begin{figure*}[t]
\centering
\resizebox{19truecm}{!}{\includegraphics[angle=-90]{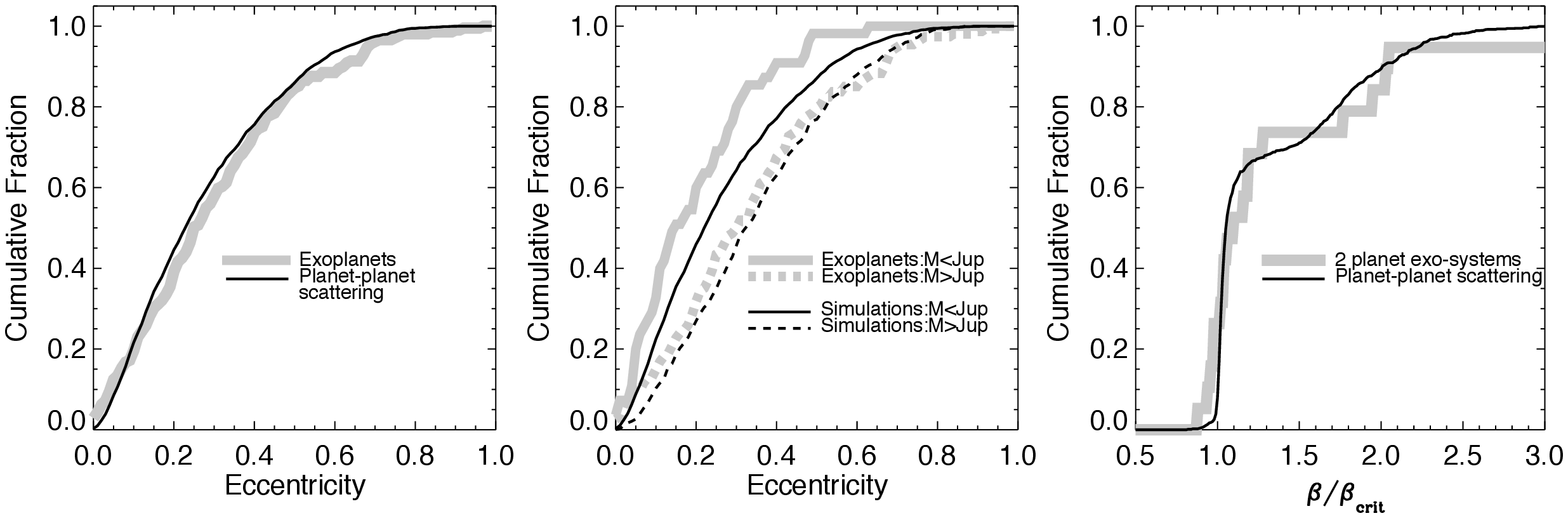}}
\caption{A comparison between observed properties of giant
exoplanets (shown in gray) and planet-planet scattering simulations
(in black).  Left: the eccentricity distribution.  Middle: the
eccentricity distribution for low-mass (dashed curves) and high-mass
(solid curves) planets.  Right: The proximity to the Hill stability
limit, measured by the quantity
$\beta/ \beta_{crit}$~\citep{barnes06b,raymond09b} where the stability limit occurs
at $\beta/ \beta_{crit}=1$ and systems with larger values of $\beta/ \beta_{crit}$ are stable.}
\label{daacfmrv_figure2}
\end{figure*}

It is relatively simple to construct a population of systems that can
reproduce the observed eccentricity distribution as well as the
mass-eccentricity correlation and also the mass distribution.  The
population consists of equal-mass high-mass systems and a diversity of
lower-mass systems that can include unequal-mass systems or equal-mass
ones.  This population also naturally reproduces the observed distribution
of two-planet systems which pile up close to the Hill stability limit (as shown 
in Figure~\ref{daacfmrv_figure2}).

We conclude this section with a note of caution.  Given the
observational uncertainties and the long list of system parameters, to
what degree can our understanding of planet-planet scattering
constrain pre-instability planetary systems?  The scattering mechanism
is robust and can reproduce observations for a range of initial
conditions.  Beyond the limitations of mass, energy and angular
momentum conservation, the details of pre-scattered systems remain
largely hidden from our view.  The only constraint that appears to
require correlated initial conditions to solve is the
mass-eccentricity correlation.

Finally, planet-planet scattering does not happen in isolation but
rather affects the other components of planetary systems.  Giant
planet instabilities are generally destructive to both inner rocky
planets~\citep{veras05,veras06,raymond11,raymond12,matsumura13} and to
outer planetesimal disks~\citep{raymond11,raymond12,raymond13}.
Additional constraints on giant planet dynamics may thus be found in a
number of places.  Recent discoveries by the Kepler Mission 
of Earth and super-Earth-size
planets relatively close to their host star reveal only a small
fraction with giant planets orbiting nearby \citep{lissauer11,ciardi13}.
Multiple systems also appear to be rather flat and stable to
planet-planet interactions \citep{lissauer11,tremaine12,johansen12,fang12b}.

Characterizing the relationship between small planets orbiting close to the
host star and more massive planets at larger separations could provide
insights into the effects of planet scattering on the formation of
inner, rocky planets.  Similarly, the abundance and mass distribution
of planets with large orbital separations \citep{malmberg11,boley12}
or free-floating planets may provide insights into the planets that
are scattered to the outskirts of planetary systems or into the galaxy
as free-floating planets \citep{veras09b,veras11,veras12,veras13}.  In
addition, the fact that planet-planet scattering perturbs both the
inner and outer parts of planetary systems may introduce a natural
correlation between the presence of debris disks and close-in low-mass
planets, as well as an anti-correlation between debris disks and
eccentric giant planets~\citep{raymond11,raymond12,raymond13}.



\bigskip
\section{\textbf{PLANETS AND PLANETESIMAL DISKS}}

This section considers the evolution of planetary orbits due to
interactions with planetesimal disks. These disks are likely to remain
intact after the gaseous portion of the disk has gone away, i.e., for
system ages greater than 3--10 Myr, and will continue to evolve in
time.  Such planetesimal disks are likely to be most effective during
the subsequent decade of time, for system ages in the range 10--100
Myr. One of the main effects of a residual planetesimal disk is to
drive planetary migration.  The subsequent changes in the orbital
elements of the planets can cause instabilities (e.g. induced by resonance
crossing or by the extraction of the planets from their original resonances), and such action can drive orbital eccentricities to larger
values. On the other hand, planetesimal disks can also damp orbital
eccentricity and thereby act to stabilize planetary systems.

We begin with a brief overview of the basic properties of planetismal
disks. Unfortunately, we cannot observe planetesimal disks directly.
Instead, we can piece together an understanding of their properties by
considering protoplanetary disks around newly formed stars 
 \citep[see the review of][]{williams11}
 and debris disks
  \citep[see the reviews of][]{zuckerman01,wyatt08}. These latter
systems represent the late stages of circumstellar disk evolution,
after the gas has been removed, either by photoevaporation or by
accretion onto the central star
(see the chapter by {\it Alexander et al.} for more details of the
photoevaporation process).

Most of the observational information that we have concerning both
types of disks is found through their spectral energy distributions
(SEDs), especially the radiation emitted at infrared wavelengths.
Since these SEDs are primarily sensitive to dust grains, rather than
the large planetesimals of interest here, much of our information is
indirect.  The defining characteristic of debris disks is their
fractional luminosity $f$, essentially the ratio of power emitted at
infrared wavelengths to the total power of the star itself. True
debris disks are defined to have $f < 10^{-2}$ 
\citep{lagrange00}, whereas systems with larger values of $f$ are considered to be
protoplanetary disks. For both types of systems, the observed fluxes
can be used to make disk mass estimates. The results show that the
disk masses decrease steadily with time. For young systems with ages
$\sim1$ Myr, the masses in solid material are typically
$M_d\sim100M_\earth$, albeit with substantial scatter about this
value. Note that this amount of solid material is not unlike that of
the Minimum Mass Solar/Extrasolar Nebula 
\citep[e.g.][]{weidenschilling77,kuchner04,chiang13}. By the time these systems reach ages of $\sim100$
Myr, in the inner part of the disk the masses have fallen to only
$\sim0.01M_\earth$ 
\citep[see Figure 3 of][]{wyatt08}. 
Note that these mass estimates correspond to the
material that is contained in small dust grains; some fraction of the
original material is thought to be locked up in larger bodies -- the
planetesimals of interest here. As a result, the total mass of the
planetesimal disk does not necessarily fall as rapidly with time as the
SEDs suggest.

At large distances from the star, instead, the decay of the mass of the dust
population is much slower, suggesting that  belts
containing tens of Earth masses can exist around Gyr-old stars \citep{booth09}.

For a given mass contained in the planetesimal disk, we expect the
surface density of solid material to initially follow a power-law form
so that $\sigma\propto{r}^{-p}$, where the index $p$ typically falls
in the approximate range $1/2 \le p \le 2$ 
\citep{cassen81}.
 The disks are initially expected to have inner edges where the
protoplanetary disks are truncated by magnetic fields, where this
boundary occurs at $r \sim 0.05$ AU. Similarly, the outer boundaries
are initially set by disk formation considerations. The angular momentum
barrier during protostellar collapse implies that disks start with
outer radii $r_d \sim 10 - 100$ AU 
\citep{cassen81,adams86}.
 Further environmental sculpting of disks
(see Section 6 below) reinforces this outer boundary. The properties
outlined here apply primarily to the starting configurations of the
planetesimal disks. As the disks evolve, and interact with planets,
the surface density must change accordingly. 


\begin{figure*}[t]
\centering
\resizebox{10truecm}{!}{\includegraphics{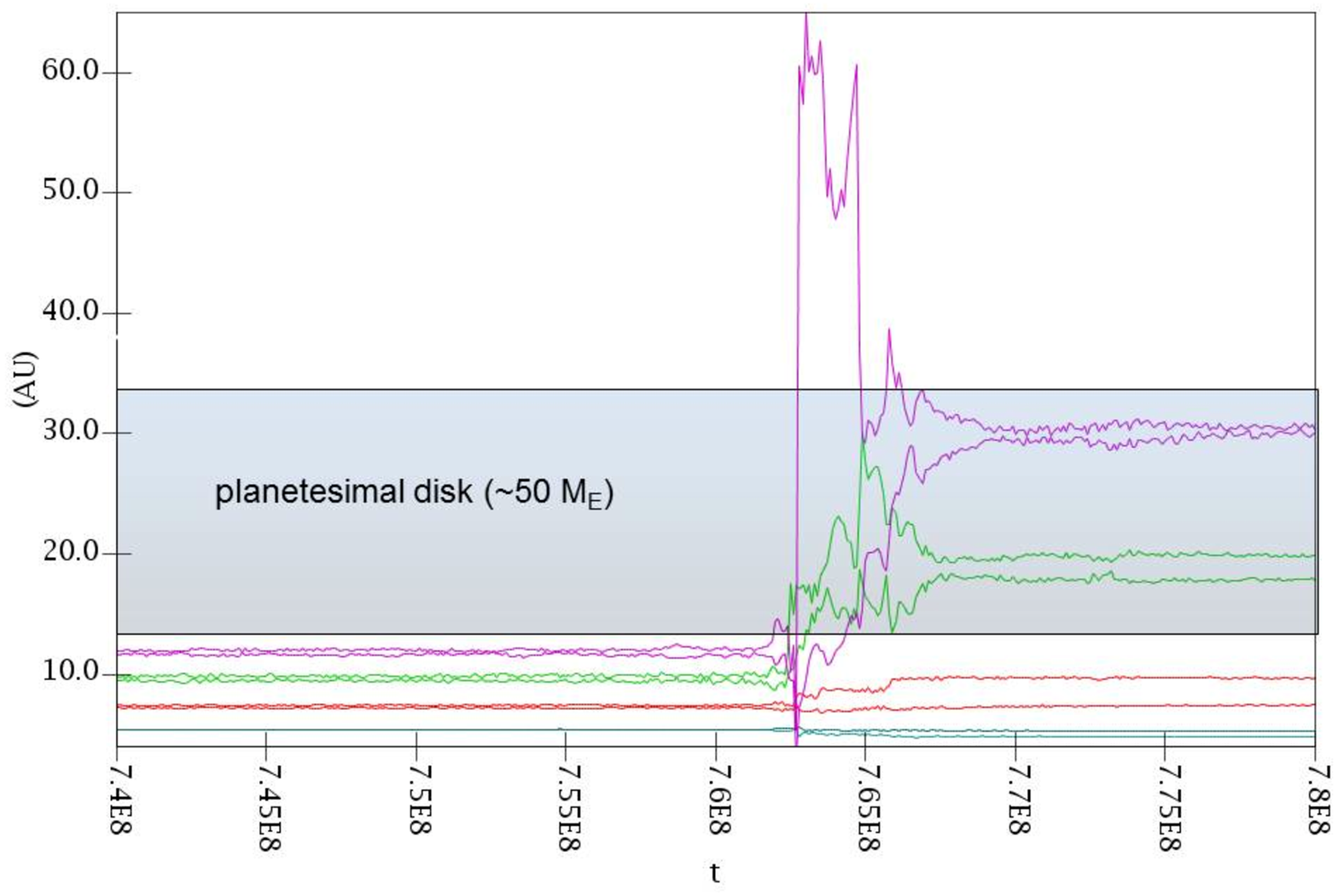}}
\caption{Orbital evolution of the 
four giant planets in our Solar System according to the Nice model.
Here, each planet is represented by two curves, denoting perihelion and aphelion distance. Hence, when the curves overlap the orbit is circular.
 The region spanned originally by the planetesimal disk of 50 Earth masses is shown as a 
 grey area. Notice that the planetary system becomes unstable at $t= 762$~Myr in this 
 simulation with the planetary orbits changing radically at this point \citep[data from][]{levison11}.}
\label{daacfmrv_figure3}
\end{figure*}

Given the properties of planetesimal disks, we now consider how they
interact with planets and drive planetary migration.  The physical
mechanism by which planetesimals lead to planet migration can be
roughly described as follows 
\citep[see, e.g.,][]{levison07}:
Within the disk, as the planetesimals gravitationally scatter off the
planet, the larger body must recoil and thereby change its trajectory.
Since the planet resides within a veritable sea of planetesimals, the
smaller bodies approach the larger body from all directions, so that
the momentum impulses felt by the planet are randomly oriented. As a
result, the orbital elements of the planet will experience a random
walk through parameter space. In particular, the semi-major axis of the
planetary orbit will undergo a random walk. If the random walk is
perfectly symmetric, then the planet has equal probability of
migrating inward or outward; nonetheless, changes will accumulate
(proportional to the square root of the number of scattering
events). In practice, however, many factors break the symmetry (e.g.,
the surface density of solids generally decreases with radius) and one
direction is preferred. 

The details of planetary migration by planetesimal disks
depend on the specific properties of the planetary system, including the
number of planets, the planet masses, their separations, the radial
extent of the disk, and of course the total mass in planetesimals. In
spite of these complications, we can identify some general principles
that guide the evolution. Some of these are outlined below: 

We first note that a disk of planetesimals containing planets often 
evolves in what can be called a ``diffusive regime'', where many small
scattering events act to make the orbital elements of both the planets
and the planetesimals undergo a random walk. As a result, the system
has the tendency to spread out. This behavior is often seen in
numerical simulations. For a system consisting of a disk of
planetesimals and analogs of the four giant planets in our Solar
System, the scattering events in general lead to Jupiter migrating inward and 
the remaining three planets migrating outward
\citep{fernandez84,hahn99,gomes04}
Similarly, in numerical simulations of
systems with two planets embedded in a disk of planetesimals, the
inner planet often migrates inward, while the outer planet migrates
outward 
\citep{levison07}.
Instead, a single planet in a planetesimal disk, in general migrates
inwards \citep{kirsh09}. This can be understood using some
considerations concerning energy conservation. For a system made of a planet
and a disk of planetesimals energy is conserved until planetesimals
are lost. Planetesimals can be lost by collisions with the planet,
collisions with the Sun or ejection onto hyperbolic orbit. In the case
where the latter is the major loss mechanism, which happens if the
planet is massive and the disk is dynamically excited, the removed
planetesimals subtract energy to the system of bodies remaining in
orbit around the star.  Consequently, most of the mass has to move
inward and the planet has to follow this trend.  Given the expected
total masses in planetesimals (see the above discussion), this mode of
migration cannot change the semimajor axis of a planetary orbit by a
large factor. In particular, this mechanism is unlikely to produce Hot
Jupiters with periods of about four days and therefore these planets have
to form by a different mechanism (migration in a disk of gas or
planet-planet scattering and tidal damping).

Finally, we note that the effects discussed above can be modified in
the early phases of evolution by the presence of a gaseous component
to the disk 
\citep{capobianco11}.
 For instance,
planetesimals scattered inwards by the planet may have their orbits
circularized by gas drag, so that they cannot be scattered again by
the planet. Consequently, the trend is that the planetesimal
population loses energy and the planet has to migrate outwards.

Planetary migration enforced by the scattering of planetesimals will
produce a back reaction on the disk. As outlined above, the disk will
tend to spread out, and some planetesimals are lost by being scattered
out of the system. Both of these effects reduce the surface density in
planetesimals. In addition, the planets can create gaps in the disk of
planetesimals. These processes are important because they are
potentially observable with the next generation of interferometers
(ALMA).  We note, however, that submillimeter (and millimeter wave)
observations are primarily sensitive to dust grains rather than
planetesimals themselves. The stirring of the planetesimal disk by
planets can lead to planetesimal collisions and dust production,
thereby allowing these processes to be observed.


Since planetesimal disks are likely to be present in most systems, it
is interesting to consider what might have occurred in the early
evolution of our own Solar System.  Here we follow the description
provided by the so-called ``Nice model". As described in Section 2, it
is expected that, at the end of the gas-disk phase, the giant planets
were in a chain of resonant (or nearly resonant) orbits, with small
eccentricities and inclinations, and narrow mutual separations. These
orbital configurations are those that are found to reach a
steady-state, and hence are used as the initial condition for the Nice
model \citep{morbidelli07}. The model also assumes that beyond the
orbit of Neptune there was a planetesimal disk, carrying cumulatively
approximately 30-50 Earth masses. A disk in this mass range is
necessary to allow the giant planets to evolve from their original,
compact configuration to the orbits they have today. More
specifically, the perturbations between the planets and this disk,
although weak, accumulated over time and eventually extracted a pair
of planets from their resonance. The breaking of the resonance lock
makes the planetary system unstable. After leaving resonance, the
planets behave as described in Sections 3 and 4, and as shown in
Figure~\ref{daacfmrv_figure3}. Their mutual close encounters act to
spread out the planetary system and to excite the orbital
eccentricities and inclinations. In particular, this model suggests
that Uranus and Neptune got scattered outwards by Jupiter and Saturn,
penetrated into the original trans-Neptunian disk and dispersed it.
The dispersal of the planetesimal disk, in turn, damped the orbital
eccentricities of Uranus and Neptune (and to a lesser extent those of
Jupiter and Saturn), so that the four giant planets eventually reached
orbits analogous to the current ones
\citep{morbidelli07,batygin10,batygin12, nesvorny12}.

In addition to the current orbits of the giant planets, the Nice model
accounts for the properties of the small body populations which, as we
described in Section 3, suggest that the structure of the solar system
experienced dramatic changes after gas dissipation. In fact, the model
has been shown to explain the structure of the Kuiper belt
\citep{levison08, batygin11}, the asteroid belt \citep{morbidelli10},
and even the origin of the Oort cloud {\citep{brasser13}. Moreover, it  
has been shown that, with reasonable assumptions concerning the
planetesimal disk, the instability of the planetary orbits could occur
after hundreds of Myr of apparent stability
\citep{tsiganis05,gomes05,levison11}; the subsequent epoch of
instability could excite the orbits of the small bodies and thereby
produce a shower of projectiles into the inner solar system that
quantitatively explains the origin of the Late Heavy Bombardment
{\citep{bottke12}.

At the present time, there are no gross characteristics of the Solar
System that are at odds with the Nice model.  Nevertheless, some
aspects of the general picture associated with the model need to be
revised or explored. For instance, the cold population (which is a
sub-population of the Kuiper belt characterized by small orbital
inclinations) probably formed in-situ \citep{parker12} instead of being
implanted from within $\sim 30$~AU as envisioned in
\citet{levison08}. Also, \citet{nesvorny12} showed that the
current orbits of the planets are better reproduced if one postulates
the existence of a fifth planet with a mass comparable to those of
Uranus and Neptune, eventually ejected from the Solar System. However,
it has not yet been shown that a 5-planet system can become unstable
late, the work of \citet{levison11} having been conducted in the
framework of a 4-planet system. These issues, however, are unlikely to
invalidate the Nice model as a whole.

It may be surprising that the current eccentricities and inclinations
of the giant planets of our solar system can be so much smaller than
those of many extra-solar planets, especially if they experienced a
similar phase of global instability (see Sections 3 and 4). There are
two main reasons for this result. First, the ``giant" planets of our
solar system have masses that are significantly smaller than those of
many extra-solar giant planets. This lower (total) mass made the
instability less violent and allowed the orbits of our giant planets
to be damped more efficiently by interactions with the planetesimal
disk \citep{raymond09a}. This trend is particularly appplicable for
Uranus and Neptune: Their eccentricities could be damped from more
than 0.5 to almost zero by the dispersal of the planetesimal disk,
which is assumed to carry about twice the sum of their masses. Second,
Jupiter and Saturn fortuitously avoided having close encounters with
each other (whereas they both had, according to the Nice model,
encounters with a Neptune-mass planet). In fact, in the simulations of
solar system instability where Jupiter and Saturn have a close
encounter with each other, Jupiter typically ends up on an orbit with
eccentricity in the range $e=0.3-0.4$ (typical of many extra-solar
planets) and recoils to ~4.5 AU, while all of the other planets are
ejected from the system. 


\bigskip
\section{\textbf{DYNAMICAL INTERACTIONS OF PLANETARY SYSTEMS WITHIN
STELLAR CLUSTERS}}

In this section we consider dynamical interactions between the
constituent members of young stellar clusters with a focus on the
consequent effects on young and forming planetary systems.

Most stars form within some type of cluster or association. In order
to quantify the resulting effects on planetary systems, we first
consider the basic properties of these cluster environments. About 10
percent of the stellar population is born within clusters that are
sufficiently robust to become open clusters, which live for 100 Myr to
1 Gyr.  The remaining 90 percent of the stellar population is born
within shorter-lived cluster systems that we call embedded clusters.
Embedded clusters become unbound and fall apart when residual gas is
ejected through the effects of stellar winds and/or supernovae.  This
dispersal occurs on a timescale of $\sim10$ Myr. 

Open clusters span a wide range of masses, or, equivalently, number of
members.  To leading order, the cluster distribution function $f_{cl} \sim
1/N^2$ over a range from $N$ = 1 (single stars) to $N=10^6$ 
\citep[this law requires combining data from different sources,
e.g.,][]{lada03,chandar99}.  With this
distribution, the probability that a star is born within a cluster of
size $N$ scales as $P = N f_{cl} \sim 1/N$, so that the cumulative
probability $\propto \log N$.  In other words, stars are equally
likely to be born within clusters in each decade of stellar membership
size $N$. For the lower end of the range, clusters have radii of order
$R$ = 1pc, and the cluster radius scales as $R \approx 1 {\rm pc}
(N/300)^{1/2}$, so that the clusters have (approximately) constant
column density 
\citep{lada03,adams06}.
For the upper end of the range, this law tends to saturate, so that
the more-massive clusters are denser than expected from this law. 
Nonetheless, a typical mean density is only about 100 stars/pc$^3$.
Clusters having typical masses and radii have velocity dispersions
$\sim 1$ km/s.

Dynamical interactions within clusters are often subject to
gravitational focussing. In rough terms, focusing becomes important
when the encounter distance is small enough that the speed of one body
is affected by the potential well of the other body. For a typical
velocity dispersion of 1 km/s, and for solar-type stars, this critical
distance is about 1000 AU. As outlined below, this distance is
comparable to the closest expected encounter distance for typical
clusters. As a result, gravitational focusing is important -- but not
dominant -- in these systems.

The timescale for a given star to undergo a close encounter 
(where gravitational focusing is important) with another
star  within a distance $r_{min}$ can be approximated by 
\citep{binney87}
\begin{eqnarray}
\label{daacfmrv_equation8}
\tau_{enc}  \simeq  3.3 \times 10^{7} {\rm yr} \left( {100 \ {\rm pc}^{-3} \over
n } \right) \left( { v_\infty \over 1 \ {\rm km/s} } \right) \nonumber \\
\times \left( { 10^3 \, {\rm AU} \over r_{min} } \right) \left( { {\rm
M}_\odot \over M} \right) 
\end{eqnarray}
Here $n$ is the stellar number density in the cluster, $v_{\infty}$ is
the mean relative speed at infinity of the stars in the cluster,
$r_{min}$ is the encounter distance, and $M$ is the total
mass of the stars involved in the encounter.  The effect of
gravitational focussing is included in the above equation.


The estimate suggested above is verified by numerical N-body
calculations, which determine a distribution of close encounters
\citep[e.g.,][]{adams06,malmberg07b,proszkow09}.
These studies show that the
distribution has a power-law form so that the rate $\Gamma$ at which a
given star encounters other cluster members at a distance of closest
approach less than $b$ has the form $\Gamma=\Gamma_0 (b/b_0)^\gamma$, 
where $(\Gamma_0,\gamma)$ are constants and $b_0$ is a fiducial
value. The index $\gamma < 2$ is due to gravitational focusing. Typical
encounter rates are shown in Figure~\ref{daacfmrv_figure4} for clusters with
$N$ = 100, 300, and 1000 members. With this power-law form for the
distribution, the expectation value for the closest encounter
experienced over a 10 Myr time span is about $\langle{b}\rangle$
$\approx$ 1000 AU. The initial conditions in a cluster can have an
important effect on the subsequent encounter rates. In particular,
sub-virial clusters which contain significant substructure (i.e.,
lumps) will have higher encounter rates
\citep[][see also Figure \ref{daacfmrv_figure4}]{allison09,parker12}.

\begin{figure}[t]
\centering
\resizebox{8truecm}{!}{\includegraphics{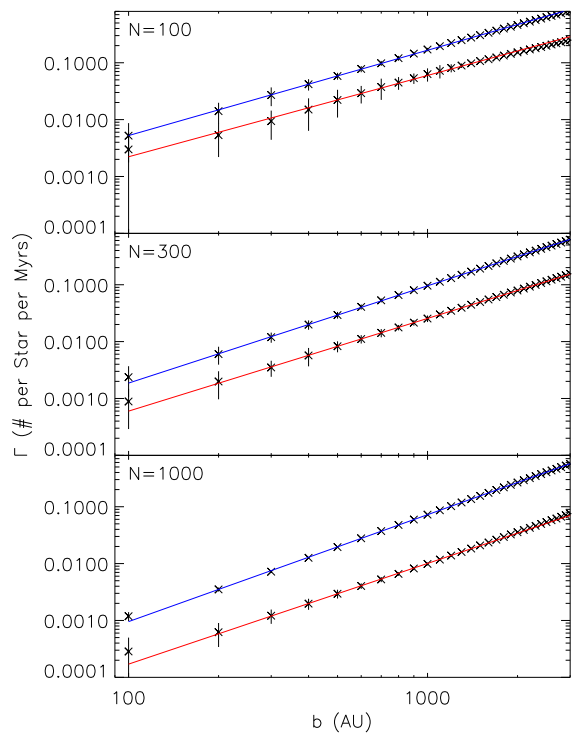}}
\caption{Distribution of closest approaches for the solar systems 
in young embedded clusters.  Each panel shows the distribution of
closest approaches, plotted versus approach distance $b$, for clusters 
with both virial (bottom) and sub-virial (top) starting conditions. 
Results are shown for clusters with $N$ = 100, 300, and 1000 members,
as labeled. The error bars shown represent the standard deviation over
the compilations (Figure 5 of {\em Adams et al.}  2006,  reproduced by permission
of the AAS).}
\label{daacfmrv_figure4}
\end{figure}

Planetary systems are affected by passing stars and binaries 
in a variety of ways
\citep[e.g.,][]{laughlin98,adams01a,bonnell01,davies01,adams06, 
malmberg07b,malmberg09,spurzem09,malmberg11,hao13}.
Sufficiently close encounters 
can eject planets, although the cross sections for direct ejections
are relatively small and can be written in the form
\citep[from][]{adams06}:
\begin{equation}
\Sigma_{ej} \approx 1350 ({\rm AU})^2 
\left( {M_\ast \over 1 {\rm M}_\odot} \right)^{-1/2} 
\left( {a_p \over 1 {\rm AU}} \right) 
\label{daacfmrv_equation9}
\end{equation}
where $M_\ast$ is the mass of the planet-hosting star and $a_p$ is the
(starting) semimajor axis of the planet. Note that the cross section
scales as one power of $a_p$ (instead of two) due to gravitational
focusing. 

Note that Equation (\ref{daacfmrv_equation9}) provides the cross section for
direct ejection, where the planet in question leaves its star
immediately after (or during) the encounter. Another class of
encounters leads to indirect ejection. In this latter case, the fly-by
event perturbs the orbits of planets in a multiple planet system, and
planetary interactions later lead to the ejection of a planet. 
Typical instability timescales lie in the range 1 -- 100 Myr
 \citep[e.g.,][see their Figure 7]{malmberg11},  although a much
wider range is possible.  On a similar note, ejection of Earth from
our own Solar System is more likely to occur indirectly through
perturbations of Jupiter's orbit (so that Jupiter eventually drives
the ejection of Earth), rather than via direct ejection from a passing
star
\citep{laughlin00}.
Planetary systems residing in wide stellar binaries in the field of the Galaxy 
are also vulnerable to external perturbations. Passing stars and the Galactic
tidal field can change the stellar orbits of wide binaries, making them eccentric,
leading to strong interactions with planetary systems \citep{kaib13}.

Although direct planetary ejections due to stellar encounters are
relatively rare, such interactions can nonetheless perturb planetary
orbits, i.e., these encounters can change the orbital elements of
planets.  Both the eccentricity and the inclination angles can be
perturbed substantially, and changes in these quantities are well
correlated
 \citep{adams01a}
 As one benchmark, the
cross section for doubling the eccentricity of Neptune in our solar
system is about $\Sigma \approx$ (400 AU)$^2$; the cross section for
increasing the spread of inclination angles in our solar system to 3.5
degrees has a similar value. Note that these cross sections are much
larger than the geometric cross sections of the Solar System. The
semi-major axes are also altered, but generally suffer smaller changes
in a relative sense, i.e., $(\Delta a)/a \ll (\Delta e)/e$.

Some time after fly-by encounters, the orbital elements of planetary
systems can be altered significantly due to the subsequent
planet-planet interactions 
\citep[see Figure 10 from][]{malmberg11}.
When a planetary system becomes unstable, planets may be
ejected via scattering with other planets. However, planets are often
ejected as the result of several (many tens to one hundred) scattering
events, with each scattering event making the planet slightly less
bound to the host star. As a result, snapshots of planetary systems
taken some time after the initial fly-by sometimes reveal planets on
much wider (less bound) orbits with separations in excess of 100
AU. In Figure \ref{daacfmrv_figure5} we plot the fraction of
post-fly-by systems containing planets on orbits with semi-major axes
$a > 100$ AU as a function of time after the fly-by encounters.  This
trend is found both in systems made unstable by fly-bys and those that
become unstable without external influence, as described earlier in 
Section 3 
\citep{scharf09,veras09a}.

Planets on such wide orbits should be
detectable via direct imaging campaigns; thus the fraction of stars
possessing them will place limits on the population of unstable
planetary systems.

\begin{figure}[t]
\centering
\resizebox{8truecm}{!}{\includegraphics{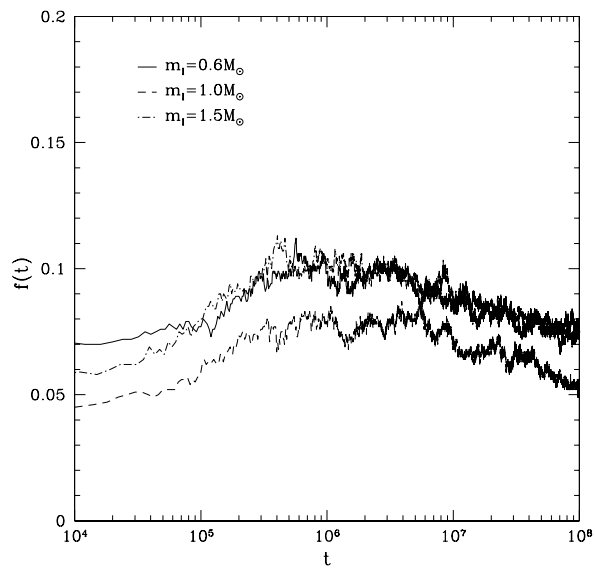}}
\caption{The fraction of solar systems $f(t)$ containing planets
with semi-major axes greater than 100 AU, plotted here as a function 
of time $t$ after a close encounter for intruder stars of
masses 0.6, 1.0 and 1.5 M$_\odot$. These simulations of fly-bys 
involve the four gas giants of the solar system with $r_{\rm min}<100$
AU (Figure 11 of {\em Malmberg et al.} 2011, reproduced 
by permission of RAS). }
\label{daacfmrv_figure5}
\end{figure}

During fly-by encounters, intruding stars can also pick up planets
from the planetary system 
\citep[see Figure 12 of][]{malmberg11}.
Clusters thus provide rich environments that can shape the
planetary systems forming within them. In particular, if the intruding
star already possesses its own planetary system, the addition of the
extra planet may destabilize the system. 

By combining encounter rates for planetary systems in clusters with
cross sections that describe the various channels of disruption, we
can estimate the probability (equivalently, the perturbation rate)
that a planetary system will suffer, e.g., the ejection of at least
one planet over a given time interval 
\citep[see][]{laughlin98,malmberg11}.
 By combining the cross section
for planetary ejection with the encounter histories found through
N-body simulations of stellar clusters, one finds that planets will be
ejected in approximately five to ten percent of planetary systems in
long-lived clusters 
\citep{malmberg11}.
The number of planets ejected directly during
fly-bys will be somewhat lower: For example, only a few planets are
expected to be ejected per (young) embedded cluster. As a result,
large numbers of free-floating planets in young clusters point to
other mechanisms for ejection, most likely planet-planet interactions
in young planetary systems
\citep{moorhead05,chatterjee08}.
 One should also note that dynamical
mechanisms are unlikely to be able to explain the population of
free-floating planets as inferred by micro-lensing observations 
\citep{veras12}.

Encounters involving binary stars also play an important role in the
evolution of planetary systems residing in stellar clusters.  A star
that hosts a planetary system may exchange into a (wide) binary. If
its orbit is sufficiently inclined, the stellar companion can affect
planetary orbits via the Kozai mechanism (see also Section 7). These
interactions force planets onto eccentric orbits that can cross the
orbits of other planets, and can thereby result in strong planetary
scatterings
 \citep{malmberg07a}.
 The rate of such encounters
depends on the binary population, but Kozai-induced scattering may
account for the destruction of at least a few percent of planetary
systems in clusters 
\citep{malmberg07b}. 

For completeness, we note that dynamical interactions can also affect
circumstellar disks, prior to the formation of planetary systems. In
this earlier phase of evolution, the disks are subject to truncation
by passing stars. As a general rule, the disks are truncated to about
one third of the distance of closest approach
 \citep{ostriker94,heller95};
 with the expected distribution of interaction
distances, we expect disks to be typically truncated to about 300 AU through
this process, although closer encounters
will lead to smaller disks around a subset of stars.
  Since most planet formation takes place at radii smaller
than 300 AU, these interactions have only a modest effect. In a similar
vein, circumstellar disks can sweep up ambient gas in the clusters
through Bondi-Hoyle accretion. Under favorable conditions, a disk can
gain a mass equivalent to the minimum mass solar nebula through this
mechanism 
\citep{throop08}.

The orientations of circumstellar disks can also be altered by 
later accretion of material on the disks and by  interactions with other
stars within clustered environments \citep{bate10}. Alternatively,
interactions with a companion star within a binary can also re-orient 
a disk \citep{batygin12b}.  Both processes represent an alternative
to the dynamical processes described in Sections 7 and 8  as a way to produce
hot Jupiters on highly-inclined orbits.

Since clusters have significant effects on the solar systems forming
within them, and since our own Solar System is likely to have formed
within a cluster, one can use these ideas to constrain the birth
environment of the Sun
 \citep{adams10}.  Our Solar System has been only moderately perturbed
 via dynamical interactions, which implies that our birth cluster was
 not overly destructive. On the other hand, a close encounter with
 another star (or binary) may be necessary to explain the observed
 edge of the Kuiper belt and the orbit of the dwarf planet Sedna
 \citep{kobayashi01,kenyon04,morbidelli04}, and the need for such an
 encounter implies an interactive environment.  Note that the expected
 timescale for an encounter (about 2000 yr) is much longer than the
 orbital period at the edge of the Kuiper belt (350 yr), so that the
 edge can become well-defined.  Adding to the picture, meteoritic
 evidence suggests that the early Solar System was enriched in
 short-lived radioactive isotopes by a supernova explosion
 \citep{wadhwa07},
 an AGB star 
\citep{wasserburg06}, or some combination of many supernovae and a
second generation massive star \citep{gounelle12, gounelle13}.  Taken
together, these constraints jointly imply that the birth cluster of
the Solar System was moderately large, with stellar membership size
$N=10^3-10^4$
\citep[e.g.,][]{hester04,adams10}.


\bigskip
\section{\textbf{THE LIDOV-KOZAI MECHANISM}}

The perturbing effect of Jupiter on the orbits of asteroids around the 
sun was considered by \citet{kozai62}. It was found that for sufficiently highly-inclined
orbits, the asteroid would undergo large, periodic, changes in both eccentricity
and inclination. Work by \citet{lidov62} showed that similar effects could be seen
for an artificial satellite orbiting a planet.  Here we will refer to such perturbations
as the Lidov-Kozai mechanism when also applied to perturbations of planetary orbits
due to an inclined stellar companion. Strong periodic interactions can also occur
between two planets, when one is highly-inclined.

The Lidov-Kozai mechanism is a
 possible formation channel for  hot Jupiters
\citep[as will be discussed in Section 8,][]{fabrycky07,nagasawa08}.
Concurrently, the Lidov-Kozai
mechanism has found wide applicability to other astrophysical 
problems:  binary supermassive black
holes 
\citep{blaes02};
binary minor planets (where the Sun is considered
the massive outer perturber) 
\citep{perets09,fang12};
binary millisecond pulsars
\citep{gopakumar09};
stellar disc-induced Lidov-Kozai oscillations in the Galactic
center
\citep{chang09};
binary white dwarfs or binary neutron 
stars
\citep{thompson11};
 and evolving triple star systems with mass loss 
 \citep{shappee13}.

\begin{figure}[t]
\centering
\includegraphics[height=90mm,width=85mm]{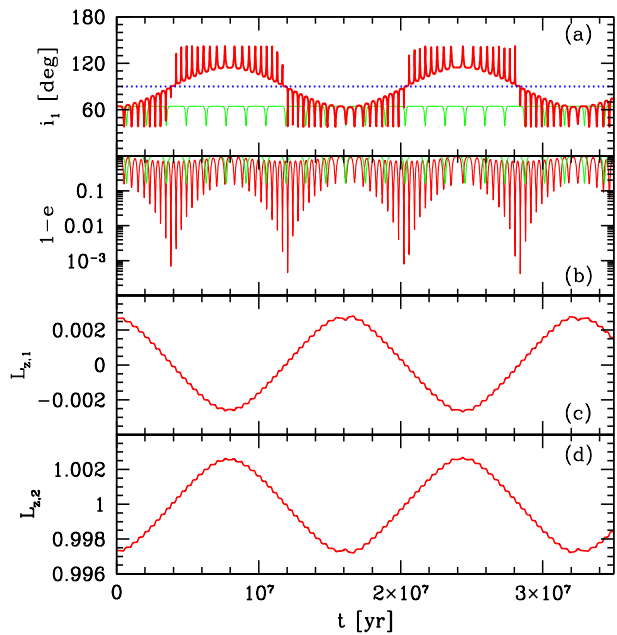}
\caption{Inclination flipping 
due to the Lidov-Kozai mechanism when the disturbing function is truncated
to octupole order (upper curve on panel a and lower curve on panel b) 
versus quadrupole order (lower curve on panel a and upper curve on panel b).
The inner binary consists of a
$1 M_{\odot}$ star and a $1 M_J$ planet separated by 6 au with $e_{\rm in} = 0.001$, and 
the outer body is a brown dwarf of $40 M_J$ at a distance of 100 au 
from the center of mass of the inner binary with $e_{\rm out} = 0.6$.
The bottom two plots display the normalized vertical components
of the inner and outer orbit angular momentum  
\citep[Figure 1 of][reprinted by permission from Macmillan Publishers Ltd: Nature, 473, 187-189, copyright 2011]{naoz11}.}
\label{daacfmrv_figure6}
\end{figure}

\begin{figure}[t]
\centering
\includegraphics[height=45mm,width=85mm]{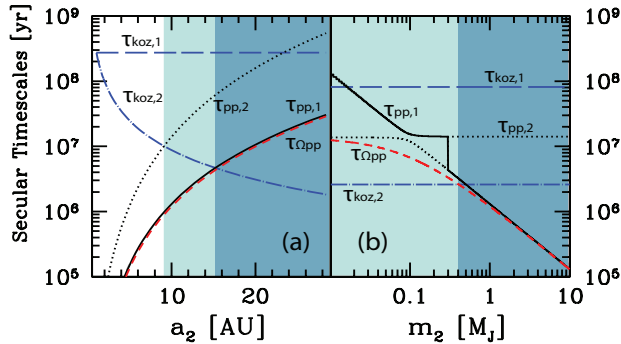}
\caption{
Comparison of timescales and regimes of motion for a four-body
system consisting of two planets secularly orbiting one star
of a wide binary.  The binary separation and eccentricity 
are fixed at 500 AU and 0.5 for two 1M$_{\odot}$ stars, and 
the inner $0.3 M_J$ planet resides 1 au away from its parent star.
Subscripts of ``1'' and ``2'' refer to the
inner and outer planet, ``koz'' the Lidov-Kozai characteristic 
timescale due to the distant star, and ``pp'' timescales due
to Laplace-Lagrange secular theory.  The unshaded, lightly-shaded
and darkly-shaded regions refer to where Lidov-Kozai cycles on the
outer planet are suppressed, the planetary orbits precess in concert, 
and the inner planet's eccentricity grows chaotically
\citep[Figure 14 of ][reproduced by permission of the AAS]{takeda08}.} 
\label{daacfmrv_figure7}
\end{figure}

Lidov-Kozai evolution is approximated analytically
by applying Lagrange's Planetary Equations to a 
truncated and averaged form of the 
disturbing function 
\citep[][Chapter 9]{valtonen06}. 
The truncation is justified because of the hierarchical ordering of the mutual
distances of the three bodies.  The averaging
occurs twice, over a longitude or anomaly 
of both the lightest body and the outermost body.  

Traditionally, the disturbing function is secular, meaning 
that the planetary semimajor axes remain fixed and the system
is free from the influence of mean motion resonances.  
The term ``Lidov-Kozai resonance'' refers
simply to the Lidov-Kozai mechanism, which includes 
large and periodic secular eccentricity 
and inclination variations.  
Confusingly, an alteration of this mechanism
that introduced non-secular contributions
\citep{kozai85}
 has been referred to as
a Lidov-Kozai resonance {\it within} a mean motion resonance.
This formalism has been utilized to help model
dynamics in the Kuiper Belt
 \citep{gallardo06,wan07,gallardo12}.
 The Lidov-Kozai mechanism can also be modified by including
 the effects of star-planet tides (see Section 8) resulting
 in significant changes in the semi-major axis of a planet
 (a process sometimes known as ``Lidov-Kozai migration'' )
\citep{wu03}.


The complete analytic solution to the secular equations
resulting from the truncated and averaged disturbing function
may be expressed in terms of elliptic functions 
\citep{vashkovyak99}.
  However, more commonly an approximate solution
is found for small initial values of eccentricities ($e$)
(but where initial values of inclination ($i$) are sufficiently large
for Lidov-Kozai cycles to occur) by truncating
the disturbing function to quadrupole order in the mutual
distances between the bodies.
This solution demonstrates that the argument of pericenter 
oscillates around $90^{\circ}$ or $270^{\circ}$, a 
dynamical signature of the Lidov-Kozai mechanism.  The solution
also yields a useful relation: 
$\sqrt{1-e^2}\cos{i} \approx {\rm constant}$
subject to $i > {\rm arcsin}(2/5)$ and
$e < \sqrt{1 - (5/3) \cos^2(i)}$.  This relation
demonstrates the interplay between eccentricity and inclination
 due to angular momentum transfer.
The period of the eccentricity and inclination oscillations 
for most observable configurations
lie well within a main-sequence star lifetime, thus at least 
thousands of such oscillations may occur before the star 
evolves.  The period
is of the order of 
\citep{kiseleva98}:

\begin{equation}
\tau_{kozai} = \frac{2 P_{out}^2}{3 \pi P_{ in}}
\frac{M_1 + M_2 + M_3}{M_3}
\left(1 - e_{out}^2 \right)^{3/2}
\label{daacfmrv_equation10}
\end{equation}

\noindent{where} $M_1$ and $M_2$ are the masses of the
innermost two bodies orbiting each other with period $P_{\rm in}$
and $M_3$ is the mass of the outermost body orbiting the inner binary with
period $P_{\rm out}$ and eccentricity $e_{\rm out}$.

Recent work has demonstrated that the approximations employed
above fail to reproduce important aspects of the true motion,
which can be modeled with 3-body numerical simulations.
By instead retaining the octupole term in the disturbing function, 
\citet{ford00b}
 derived more accurate, albeit complex,
evolution equations for the true motion.  Subsequent relaxation
of the assumption of small initial eccentricities has
allowed for a wider region of phase space of the true motion
to be reproduced by the Lidov-Kozai mechanism
\citep{katz11,lithwick11,naoz11,libert12}.

One outstanding consequence of retaining the octupole term
is that the effect may flip a planet's orbital evolution from prograde
to retrograde, and consequently may explain observations.
The projected angle between stellar rotation and  planetary orbital angular
momentum has been measured for tens of hot Jupiters with the Rossiter-Mclaughlin
effect 
\citep{triaud10}. These observations show us that some 20\% of hot jupiters
most probably have retrograde orbits whilst 50\% or so are aligned \citep{albrecht12}.
 In at least one case
  \citep{winn09}
the true angle is at least $86^{\circ}$, helping to reinforce 
indications of a subset of planets orbiting in a retrograde fashion with inclinations
above $90^\circ$
 \citep{albrecht12}.
\citet{naoz11}
 demonstrated how the Lidov-Kozai mechanism can
produce these orbits\footnote{They also discovered an important error in the 
original derivation of the truncated, averaged disturbing
function
 \citep{kozai62}
 which did not conserve angular momentum
due to an erroneous assumption about the longitudes of ascending nodes.}.  
 Figure~\ref{daacfmrv_figure6} illustrates
how this inclination ``flipping'' can occur naturally in a three-body
system, and the consequences of truncating the disturbing function
to quadrupole order. In panel (a) and (b) of this figure, one sees that when the
disturbing function is truncated at the quadrupole term, then the spread of 
inclinations and eccentricities are small, whereas when going to octupole order,
the orbit flips (ie inclinations above $90^\circ$) and eccentricities reach
to values very close to unity.

Lidov-Kozai oscillations do not work in isolation.
Bodies are not point masses: effects of tides,
stellar oblateness and general relativity may
play crucial roles in the evolution.  These contributions
have been detailed by 
\citet{fabrycky07}, \citet{chang09}, \citet{veras10} and \citet{beust12}.

A major consequence of these effects, in particular
tides, is that planets formed beyond the snow line
may become hot Jupiters through
orbital shrinkage and tidal circularization [see Section 8].

The Lidov-Kozai mechanism may also be affected by the 
presence of a nascent protoplanetary disc, and hence play a crucial 
role during the formation of planets around one component of a binary system.
If the secondary star is sufficiently inclined to and separated from
the orbital plane of the primary, then the Lidov-Kozai
mechanism might ``turn on'', exciting the eccentricity of planetesimals
and diminishing prospects for planet formation 
\citep{marzari07,fragner11}.
However, the inclusion of the effects of gas drag
 \citep{xie11}
and protoplanetary disc self-gravity
 \citep{batygin11}
assuage the destructive effect of Lidov-Kozai oscillations,
helping to provide favorable conditions for planetary growth.

Lidov-Kozai-like oscillations are also active in systems
with more than three bodies, which is the focus
of the remainder of this section.  Examples
include quadruple star systems 
\citep{beust06},
triple star systems with one planet 
\citep{marzari07}
and multiple-planet systems with or without
additional stellar companions.  

Given the likelihood of multiple exoplanets in binary
systems,
 how the Lidov-Kozai mechanism
operates in these systems is of particular interest.
In particular, eccentricity and inclination oscillations
produced by a wide-binary stellar companion can
induce planet-planet scattering, leading to dynamical
instability
 \citep{innanen97}.
\citet{malmberg07a}
demonstrate that
this type of instability can result in planet
stripping in the stellar birth cluster, where
binaries are formed and disrupted at high
inclinations.

Alternatively, multiple-planet systems within
a wide binary may remain stable due to Lidov-Kozai
oscillations.  Understanding the conditions
in which stability may occur and the consequences
for the orbital system evolution can help explain
current observations.
 \citet{takeda08}
outline how to achieve this characterization by 
considering the secular
evolution of a two-planet system
in a wide binary, such that the mutual planet-planet
interactions produce no change in semimajor axis.
These planet-planet interactions may be coupled
analytically to Lidov-Kozai oscillations because
the latter are usually considered to result from
secular evolution.

 \citet{takeda08}
 compare the period of Lidov-Kozai oscillations with the period
the oscillations produced by Laplace-Lagrange
secular interactions (see Section 2), as
shown in  Figure~\ref{daacfmrv_figure7}.  The figure provides
an example of where Lidov-Kozai oscillations
dominate or become suppressed (by comparing the curves), 
and the character of the resulting dynamical
evolution (identified by the shaded regions).
Because the Lidov-Kozai oscillation timescale increases
with binary separation (see Equation \ref{daacfmrv_equation10}),
  Lidov-Kozai oscillations are  less likely to have 
 an important effect on multi-planet evolution contained in wider stellar binaries.
However, for wide-enough binaries, Galactic tides can
cause close pericenter passages every few Gyr, perhaps
explaining the difference in the observed eccentricity
distribution of the population of giant planets
in close binaries versus those in wide binaries 
\citep{kaib13}.


\bigskip
\section{\textbf{DYNAMICAL ORIGIN OF HOT JUPITERS}}

The first exoplanet that was confirmed to orbit a main sequence star, 
51 Pegasi b 
\citep{mayor95}
 became a prototype for a class of 
exoplanets known as hot Jupiters. The transiting subset of the 
hot Jupiters (here defined as $a \le 0.1$ AU and $M_p \sin i = 0.25-20 M_J$) 
provide unique constraints on planetary evolution, and allow observational 
studies of atmospheric phenomena that are currently not possible for more 
distant planets.
Identifying the dynamical origin of hot-Jupiters is a long-standing problem 
and is the topic of this section.

\begin{figure}[t]
\centering
\includegraphics[height=100mm,width=85mm]{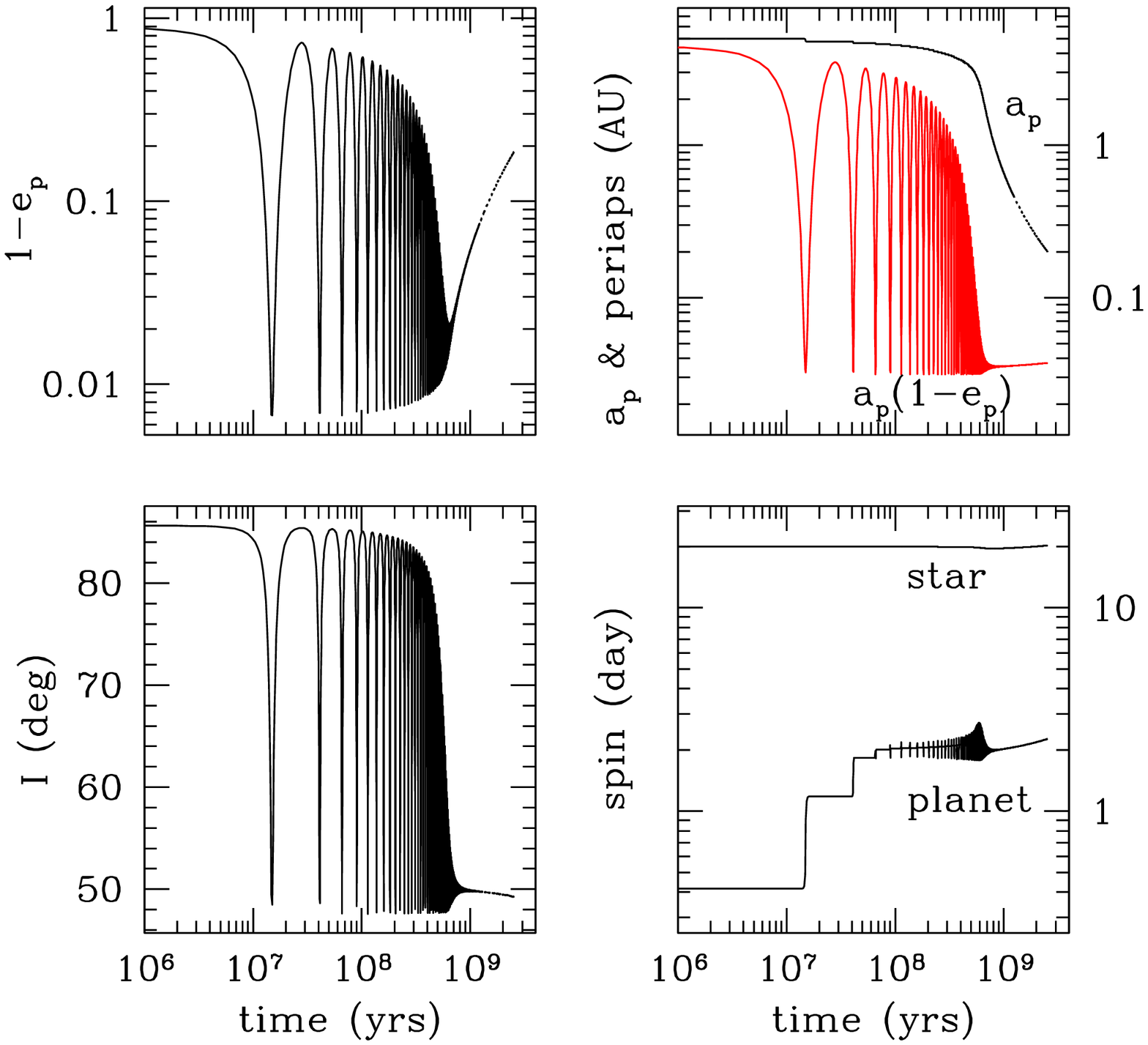}
\caption{Hot Jupiter production by Lidov-Kozai
forcing from a stellar binary companion.  A $7.80 M_J$ planet with $a_p = 5.0$ AU,
$e_p(t=0) = 0.1$, $I_p(t=0) = 85.6^{\circ}$, and $\omega_p(t=0) = 45^{\circ}$  
is evolving under the influence of two $1.1 M_{\odot}$ stars separated by $1000$ AU
on a $e_B = 0.5$ orbit
\citep[Figure1 of ][reproduced by permission of AAS]{wu03}.}
\label{binary}
\label{daacfmrv_figure8}
\end{figure}

\begin{figure}[t]
\centering
\includegraphics[height=120mm,width=85mm]{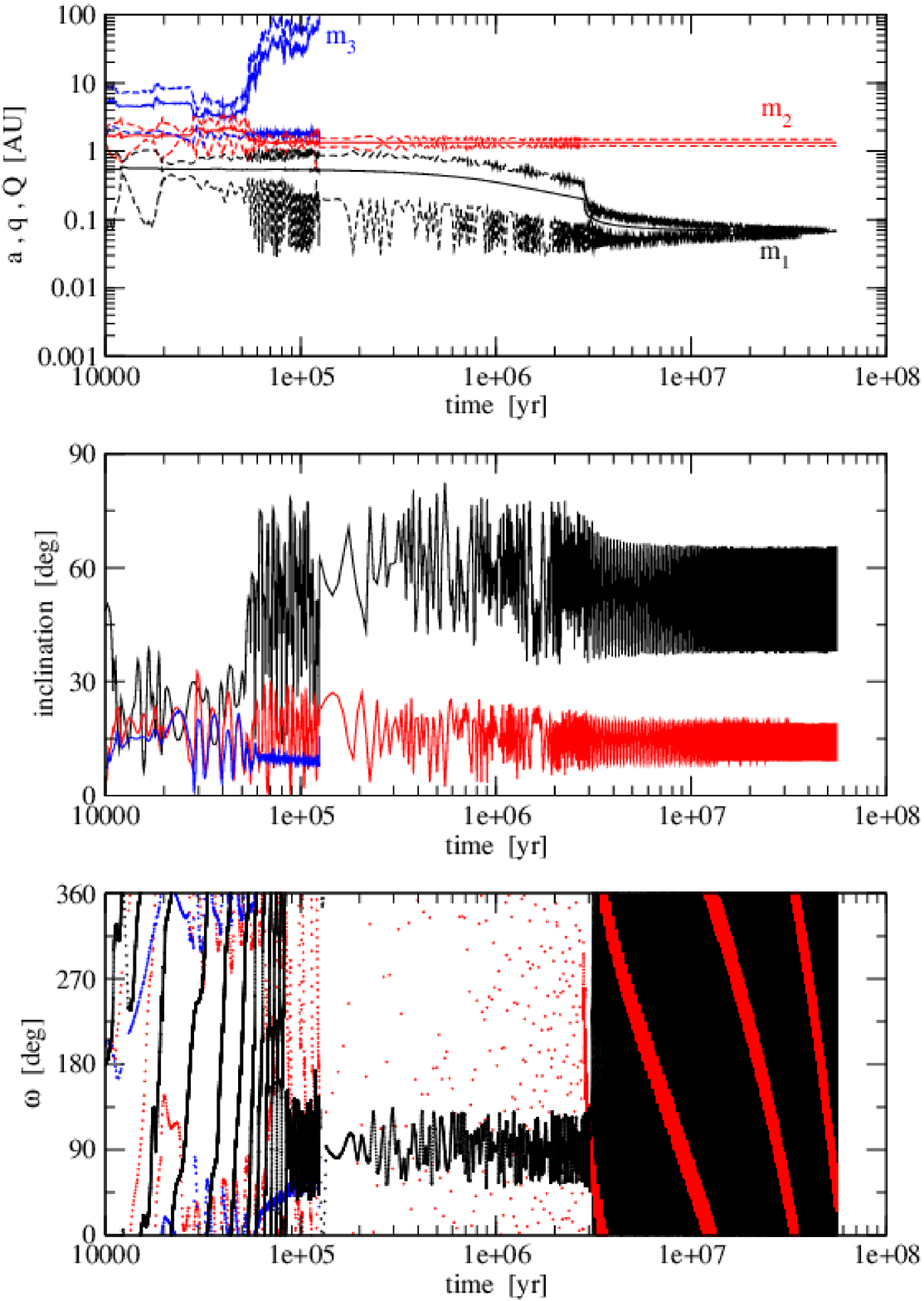}
\caption{Hot Jupiter production by Lidov-Kozai
forcing from multiple-planet interactions.  The inner, middle and outer planets have masses
of $1M_J$, $2M_J$ and $1M_J$ and initially nearly circular ($e < 0.1$) and coplanar ($I < 1^{\circ}$) orbits
 \citep[Figure 11 of ][reproduced by permission of AAS]{beauge12}.}
\label{daacfmrv_figure9}
\end{figure}

It is generally accepted that hot Jupiters cannot form {\em in situ} 
\citep{bodenheimer00}.
If true, then their origin requires either migration 
through a massive, and presumably gaseous, disk or dynamical interactions 
involving multiple stellar or planetary bodies. 
Although both possibilities remain open, recent observations
\citep[e.g.,][]{winn10}
indicate that roughly one fourth
of hot Jupiter orbits are substantially misaligned with respect to the
stellar rotation axis. These systems (and perhaps others) are naturally
explained by dynamical processes, which are the focus of this
section.
The relevant dynamics may involve:
\begin{itemize}
\item Lidov-Kozai evolution of a one-planet system perturbed by a 
binary stellar companion 
\item Lidov-Kozai evolution in a multiple-planet system 
\item Scattering of multiple planets or secular evolution unrelated to the 
Kozai resonance
\item Secular chaos
\end{itemize}

\begin{figure*}[t]
\centering
\includegraphics[height=140mm,angle=-90]{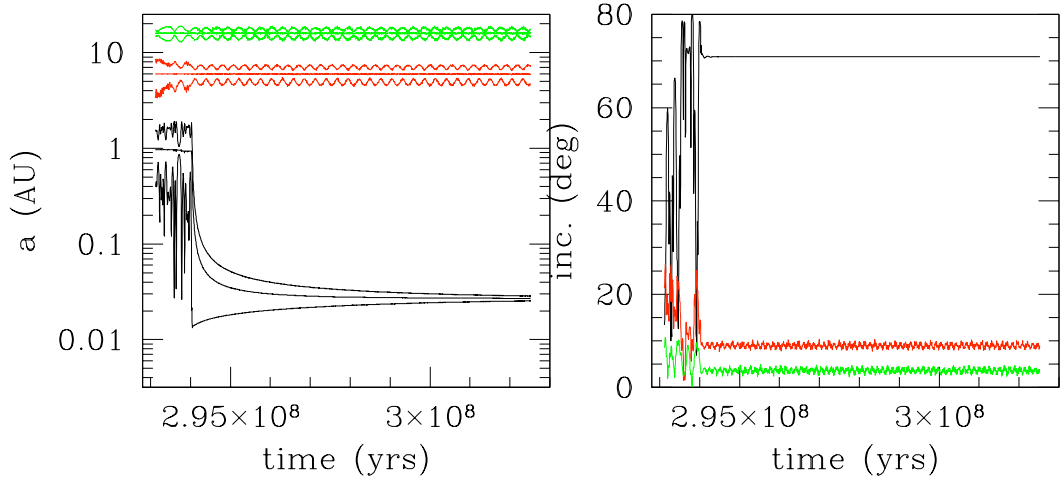}
\caption{Hot Jupiter production by multiple planet
evolution with no Lidov-Kozai cycles. The inner, middle and outer planets have masses
of $0.5M_J$, $1.0M_J$ and $1.5M_J$, and initial semimajor axes of $1$ AU, $6$ AU and $16$ AU, 
eccentricities of $0.066$, $0.188$ and $0.334$ and inclinations of $4.5^{\circ}$, $19.9^{\circ}$
and $7.9^{\circ}$.   This choice places the AMD primarily in the outer planets
\citep[Figure 2 of ][reproduced by permission of AAS]{wu11}.}
\label{daacfmrv_figure10}
\end{figure*}

All of these processes are likely to occur at some level, so the main open question is 
their relative contribution to forming the observed hot Jupiter population.  
In every case, tidal interactions --- which are inevitable for planets with 
the orbital period of hot Jupiters --- are required in order to 
shrink and circularize the orbit
 \citep{ivanov04,guillochon11}.   
Tides raised on the star are expected to dominate orbital decay, while tides raised on the planet are likely to 
dominate circularization.  However, tides on both bodies can contribute significantly.  
Quantitative comparisons between theoretical models and observations are 
limited by considerable uncertainties in tidal physics and planetary structure, as well as observational biases 
in determining the orbital period distribution and the eccentricity of nearly circular orbits 
\citep{zakamska11,gaidos13}.
%
%

A mechanism for generating high eccentricities, and tidal damping, are minimal 
ingredients for the dynamical formation of hot Jupiters. Several other processes, 
however, including general relativity, oblateness
 \citep{correia12}
  and tides, can lead to 
precession of short-period orbits. These processes need to be included in 
models of hot Jupiter formation via secular dynamical effects. {\em Wu and Murray} 
(2003) quote formulae for the precession rates,
\begin{eqnarray} 
 \dot{\omega}_{GR} & = & 3n \frac{GM_*}{a_p c^2 (1-e_p^2)} \nonumber \\
 \dot{\omega}_{tid} & = & \frac{15}{2} n k_2 
   \frac{1 + (3/2) e_p^2 + (1/8) e_p^4}{(1-e_p^2)^5} \frac{M_*}{M_p} 
   \left( \frac{R_p}{a_p} \right)^5 \nonumber \\
 \dot{\omega}_{obl} & = & \frac{1}{2} n \frac{k_2}{(1-e_p^2)^2} 
 \left( \frac{\Omega_p}{n} \right)^2 \frac{M_*}{M_p} 
   \left( \frac{R_p}{a_p} \right)^5 .
\end{eqnarray} 
Here $a_p$, $e_p$, $M_p$, $R_p$ and $\Omega_p$ are the planet's semi-major axis, 
eccentricity, mass, radius and spin frequency, $n$ is the mean motion, and $k_2$ 
is the tidal Love number. 
Lidov-Kozai oscillations are generally suppressed if any of these precession rates exceed 
$\dot{\omega}_{kozai}$. This can easily occur for $a_p < 1 \ {\rm AU}$.  Eccentricity can be
enhanced when $\dot{\omega}_{GR}\simeq-\dot{\omega}_{kozai}$ 
\citep{ford00b}.

Figure~\ref{daacfmrv_figure8}, from 
 \citet{wu03},
 illustrates how hot Jupiters 
can form in a single planet system with an inclined stellar companion. Lidov-Kozai oscillations 
in the planetary eccentricity result in periods where the pericenter distance is close 
enough for stellar tides to shrink the orbit. As the orbit shrinks, the additional 
dynamical effects described above become more important, resulting in a decrease in 
the amplitude of the eccentricity and inclination variations. In this example, after 700~Myr, GR suppresses
the Lidov-Kozai oscillations completely. The influence of
the secondary star then becomes negligible as tidal interactions dominate. 
\citet{wu07}, using a binary population model and estimates for the radial distribution 
of massive extrasolar planets, estimated that this process could account for 10\% or more 
of the hot Jupiter population.

A combination of gravitational scattering \citep{ford06} and Lidov-Kozai oscillations can also 
lead to the production of hot Jupiters 
from multiple-planet systems around single stars.
Figure~\ref{daacfmrv_figure9} provides an example of three-planet scattering
in which the outer planet is ejected, triggering Lidov-Kozai cycles between
the other two planets.  These oscillations are not as regular
as those in Figure~\ref{daacfmrv_figure6} because in Figure~\ref{daacfmrv_figure9} 
the outermost body is comparable in mass to the middle body.  
Nevertheless,  the oscillations of the
argument of pericenter about $90^{\circ}$ is indicative of the Lidov-Kozai
mechanism at work.  By about 3 Myr, the semimajor axis of the inner planet
has shrunk to a value of 0.07 AU, after which other physical effects dictate
the future evolution of the planet.

Both the fraction of scattering systems that yield 
star-grazing planets, and the fraction of those systems that yield surviving 
hot Jupiters, are uncertain. 
\citet{nagasawa08} and \citet{nagasawa11} integrated ensembles of unstable 
three planet systems, using a model that included both gravitational and tidal forces. 
They obtained an extremely high yield ($\simeq 30$\%) of highly eccentric planets, 
that was larger than the yield found in earlier calculations that did not include 
tides \citep{chatterjee08}. One should note that these numbers do not reflect the 
expected fraction of scattering systems that would yield {\em long-lived} hot 
Jupiters, as many of the highly eccentric planets circularize into orbits with 
tidal decay times less than the main sequence lifetime. 
\citet{beauge12}, on the other hand, using a different 
tidal model, a dispersion in planet masses, and resonant initial conditions, found a 
yield of surviving hot Jupiters of approximately 10\%. Ten percent is also, roughly, 
the fraction of hot Jupiters in an unbiased sample of all massive planets with 
orbital radii less than a few AU. The efficiency of hot Jupiter production from 
scattering plus tidal circularization is thus high enough for this channel 
to contribute substantially to the population, if one assumes that scattering occurs 
in the majority of all such planetary systems.   

Hot Jupiters can originate from multi-planet dynamics without
the Lidov-Kozai effect. 
 \citet{wu11}
 present special configurations of the 4-body problem that allow
for the innermost planet in a 3-planet system to be forced
into the tidal circularization radius.  These configurations
require a significant angular momentum deficit (AMD) in 
the original planetary system. 
Figure~\ref{daacfmrv_figure10} presents an example of hot Jupiter production
without the Lidov-Kozai effect. Note that all three planets survive
the evolution and never cross orbits.  

If planet formation produces multiple planets on nearly circular, coplanar orbits, 
then secular evolution alone is not sufficient to produce
hot Jupiters. Planet-disk interactions are not currently thought to be able to 
generate significant planetary eccentricity (except for high mass planets), 
and hence the easiest route toward forming systems with a significant AMD 
appears to be an initial phase of strong planet-planet scattering. In such a 
model, hot Jupiters could either be formed early (from highly eccentric scattered planets) 
or late (from long term secular evolution among the remaining planets after scattering). 
Which channel would dominate is unclear. 
 
The observation of strongly misaligned and retrograde orbits from Rossiter-McLaughlin measurements 
provides evidence in favor 
of dynamical formation mechanisms
 \citep[e.g.,][]{winn09},
  but does not immediately 
discriminate among different dynamical scenarios. In particular, 
although pure Lidov-Kozai evolution involving a circular stellar companion 
cannot create a retrograde planet, the presence of eccentricity 
in either a stellar or planetary perturber can
 \citep{katz11,lithwick11}.
 The same is true for 
secular evolution without the Lidov-Kozai effect.

\citet{dawson12b}
 compare the three  potential origin channels described above, and 
 suggest that Lidov-Kozai
evolution in binary  stellar systems is unlikely to  dominate, but may explain $15_{-11}^{+29}\%$ of hot Jupiters,  corroborating
 \citet{naoz12}
 result of  $\approx 30\%$. These estimates rely upon the results of 
 \citet{nagasawa11} 
  that find a large
fraction of scattering  systems ($\sim 30\%$) form (at least initially) 
a hot Jupiter, a result that they attribute in substantial part to Lidov-Kozai evolution. 
Similar calculations by 
\citet{beauge12},
using a more realistic tidal model, find less efficient hot Jupiter formation driven 
primarily by scattering events.  They
suggest that higher initial planetary  multiplicity results in hot Jupiter populations in better accord with observations. 
\citet{morton11}
provide constraints on the frequency of the two
different Lidov-Kozai-induced hot Jupiter formation scenarios 
from spin-orbit data.  They find that the results depend critically
on the presence or lack of a population of aligned planetary systems.
If this population exists, multiple-planet Lidov-Kozai scattering 
 \citep[specifically from the model of ][]{nagasawa08}
is the favored formation mechanism.  Otherwise, binary-induced Lidov-Kozai evolution
\citep[specifically from the model of ][]{fabrycky07}
is the favored model.  The above results also depend upon the assumed tidal physics.  
\citet{dawson12a}
identified Kepler Object of Interest (KOI) 1474.01 as a proto-hot Jupiter based on the long
 transit duration for its orbital period.  The presence of large transit timing variations
  suggests that scattering by a more distant giant planet may explain the origin of KOI 
  1474.01's high eccentricity.  With further analysis, the abundance of transiting proto-hot
   Jupiters could provide a constraint on the timescale of the high eccentricity
    migration phase of hot Jupiter's formation.


\bigskip
\section{\textbf{SUMMARY}}

We have reviewed the long-term dynamical evolution
of planetary systems. Our key points are listed below:

\begin{enumerate}
\item {\em The giant-planet sub-system of the Solar System is stable}
although the terrestrial-planet sub-system is marginally unstable with  
a small chance of planet-planet encounters during the lifetime of the Sun.
\item {\em Planet-planet scattering}  in tighter planetary systems
can lead to close encounters between planets. The timescale before a system undergoes
such encounters is a strong function of  the separation of planets. 
\item {\em Secular interactions}  cause the redistribution of  angular momentum
amongst planets in a system. In systems with a sufficiently large angular momentum 
deficit (AMD), such redistribution can lead to  close planetary encounters.
\item {\em The outcome  of planetary close encounters } is a function of the  Safronov 
number.  Collisions dominate when the planetary surface escape speeds are
smaller than orbital speeds. Planetary scattering will be more common
when the surface escape speeds are larger than the planetary orbital speeds in a
system.
\item {\em Planets are predicted to pass through a phase of wide orbits} within
unstable planetary systems as  ejections occur only after 
several scatterings.  Imaging surveys will therefore inform us
about the frequency of unstable systems.
\item {\em The observed eccentricity distribution} is consistent with being an outcome
of planet-planet scattering in unstable systems. 
\item {\em Interactions with planetesimal disks} will cause planets to migrate
which in turn can lead to instabilities within a planetary system. This process
probably played an important role in the early history of our own Solar System.
\item {\em Aging systems may become unstable} when the host star evolves
to become a white dwarf, and loses mass,   as the
relative strength of the planet-planet interactions increase compared to the interactions
between the planets and host star.
\item {\em Fly-by encounters in stellar clusters} will occur in dense birth environments.
Such encounters may lead to the direct ejection of planets in some cases. In other
encounters, perturbations to the planetary orbits lead to instabilities on longer
timescales. The intruding star may also pick-up a planet from the system.
\item {\em Exchange into binaries} can occur in stellar clusters. Planetary systems
may be de-stabilised by the perturbing effect of the companion star through the 
Lidov-Kozai mechanism where the outer planet suffers periods
of higher eccentricity leading it to have strong encounters with other planets.
\item {\em The Lidov-Kozai mechanism} may also operate within primordial
binaries or planetary multiple systems, leading to the periodic increase 
in eccentricity of planet's orbits and planet-planet encounters in the case
of multiple-planet systems.
\item {\em The origin of  hot Jupiters} through dynamical interactions
may involve one of five
possible routes:  Lidov-Kozai evolution of a one-planet system perturbed 
by a binary stellar companion; Lidov-Kozai evolution in a multiple-planet system;
scattering of multiple planets; secular evolution unrelated to the Kozai resonance;
or the re-orientation of circumstellar disks before planets form via interaction with
a companion star, or via late infall of material or interactions or neighboring stars in
clustered birth environments.
\end{enumerate}



\noindent
\textbf{ Acknowledgments.} MBD  was supported by the Swedish Research
Council (grants 2008-4089 and 2011-3991).
PJA acknowledges support from NASA grants NNX11AE12G and NNX13AI58G, and from 
grant HST-AR-12814 awarded by the Space Telescope Science Institute, which is operated
 by the Association of Universities for Research in Astronomy, Inc., for NASA, under contact 
 NAS 5-26555.
 JC would like to thank NASA's Origins of Solar Systems program for support.
  EBF  was supported by the National Aeronautics and Space
Administration under Origins of Solar Systems grant NNX09AB35G and
grants NNX08AR04G and NNX12AF73G issued through the Kepler
Participating Scientist Program. We thank the anonymous referee for their useful comments.




\bibliographystyle{ppvi_lim6}
\bibliography{refs_all}

\end{document}